\documentclass[twocolumn]{aastex701}

\usepackage{tabularx} 
\usepackage{booktabs}
\usepackage{afterpage} 
\usepackage{threeparttable} 
\usepackage{subcaption}
\usepackage{multirow} 
\usepackage{graphicx}
\usepackage{amsmath} 
\usepackage{amssymb}
\usepackage{hyperref} 
\usepackage{rotating}
\usepackage{float}
\bibpunct{(}{)}{;}{a}{}{,}  

\graphicspath{{./}{Fig/}}

\shorttitle{Dust Reverberation Mapping of Low-Mass AGNs}
\shortauthors{Wang et al.}

\begin{document}
\title{Dust Reverberation Mapping of Low-Mass Active Galactic Nuclei}
\author[0009-0002-5955-4932]{Chi-Zhuo Wang}
\affiliation{Department of Astronomy, School of Physics, Peking University, Beijing 100871, People's Republic of China}
\affiliation{Kavli Institute for Astronomy and Astrophysics, Peking University, Beijing 100871, People's Republic of China}
\email{chizhuowang@stu.pku.edu.cn}

\author[0000-0001-8416-7059]{Hengxiao Guo}
\affiliation{Shanghai Astronomical Observatory, Chinese Academy of Sciences, 80 Nandan Road, Shanghai 200030, People's Republic of China}
\email{hengxiaoguo@gmail.com}

\author[0000-0002-7350-6913]{Xue-Bing Wu}
\affiliation{Department of Astronomy, School of Physics, Peking University, Beijing 100871, People's Republic of China}
\affiliation{Kavli Institute for Astronomy and Astrophysics, Peking University, Beijing 100871, People's Republic of China}
\email{wuxb@pku.edu.cn}

\author[0000-0002-4521-6281]{Wenwen Zuo}
\affiliation{Shanghai Astronomical Observatory, Chinese Academy of Sciences, 80 Nandan Road, Shanghai 200030, People's Republic of China}
\email{wenwenzuo@shao.ac.cn}

\author[0009-0007-1193-609X]{Wanling Liu}
\affiliation{Shanghai Astronomical Observatory, Chinese Academy of Sciences, 80 Nandan Road, Shanghai 200030, People's Republic of China}
\affiliation{University of Chinese Academy of Sciences, 19A Yuquan Road, 100049, Beijing, People's Republic of China}
\email{wanlingliu@shao.ac.cn}

\author[0000-0003-3472-4392]{Jingbo Sun}
\affiliation{Shanghai Astronomical Observatory, Chinese Academy of Sciences, 80 Nandan Road, Shanghai 200030, People's Republic of China}
\affiliation{University of Chinese Academy of Sciences, 19A Yuquan Road, 100049, Beijing, People's Republic of China}
\email{sunjingbo@shao.ac.cn}

\author[0000-0002-3742-6609]{Wenke Ren}
\affiliation{Shanghai Astronomical Observatory, Chinese Academy of Sciences, 80 Nandan Road, Shanghai 200030, People's Republic of China}
\email{wenke.ren@shao.ac.cn}

\author[0000-0002-2052-6400]{Shu Wang}
\affiliation{School of Physics \& Astronomy, University of Southampton, Southampton, SO17 1BJ, United Kingdom} 
\email{wangshu100002@gmail.com}

\correspondingauthor{\scriptsize Chi-Zhuo\,Wang,\,Hengxiao\,Guo,\,Xue-Bing\,Wu}
\email{chizhuowang@stu.pku.edu.cn, hengxiaoguo@gmail.com,\\ wuxb@pku.edu.cn}

\begin{abstract}
We perform a systematic dust reverberation mapping study of low-mass ($M_{\mathrm{BH}}\sim10^6$-$10^7 M_\odot$) active galactic nuclei (AGNs) selected from a literature-compiled catalog, using data from the Zwicky Transient Facility and the Wide-field Infrared Survey Explorer. Through quality assessments, we obtain a core sample of 65 sources with W1- and W2-band lag measurements relative to optical light curves. The W2-band lags are generally longer than the W1-band lags, with a median ratio of $\tau_{\mathrm{W2}}/\tau_{\mathrm{W1}}=1.24^{+0.33}_{-0.21}$. We further investigate the relations between dust radius and AGN luminosity ($R_{\mathrm{dust}}-L_{\mathrm{AGN}}$), using different lag measurements and luminosity proxies. Combining our sample with a higher-luminosity AGN sample, we establish unified $R_{\mathrm{dust}}-L_{\mathrm{AGN}}$ relations spanning nearly seven orders of magnitude in bolometric luminosity. The slopes are $0.47\pm0.01$ and $0.44\pm0.01$ for the W1 and W2 bands, slightly below the expectation of $0.5$ from dust radiative equilibrium. In the combined sample, the lag deviations show a negative correlation with Eddington ratio, potentially related to the self-shadowing effect of slim disks. The luminosity-dependent SED variations revealed by the SPHEREx composite spectra suggest that individual luminosity proxies may not fully trace the intrinsic dust-heating luminosity. Both effects may contribute to the shallow $R_{\mathrm{dust}}-L_{\mathrm{AGN}}$ slopes. Future joint optical and near-infrared monitoring with the Vera C. Rubin Observatory's Legacy Survey of Space and Time and the Nancy Grace Roman Space Telescope will further extend dust RM studies to lower-luminosity AGNs hosting lower-mass black holes.

\end{abstract}


\section{Introduction} \label{sec:intro}
Active galactic nuclei (AGNs) are powered by accretion onto supermassive black holes (SMBHs) and emit radiation across the electromagnetic spectrum. In the standard unification picture, each central engine is surrounded by an accretion disk, a broad-line region (BLR), and an obscuring dusty torus, whose orientation largely determines the observed spectral classification of AGNs \citep{1993ARA&A..31..473A,1995PASP..107..803U,2015ARA&A..53..365N}. The dusty torus absorbs ultraviolet and optical radiation from the accretion disk and re-emits it in the infrared, producing the characteristic infrared bump in AGN spectral energy distributions (SEDs) \citep{1978ApJ...226..550R,2008ApJ...685..160N,2010A&A...523A..27H}. The dusty torus is commonly modeled as a clumpy distribution of optically thick clouds \citep{1988ApJ...329..702K,2008ApJ...685..147N}, while high-angular-resolution infrared observations have revealed extended polar dust emission in some AGNs, possibly associated with dusty winds \citep{2013ApJ...771...87H,2017ApJ...838L..20H,2019ApJ...884..171H}. The dust in the torus is generally assumed to consist of a mixture of graphite and silicate grains, whose different optical properties and sublimation temperatures shape the infrared continuum and help determine the inner dust boundary \citep{1987ApJ...320..537B,1993ApJ...402..441L,2007A&A...476..713K,2012MNRAS.420..526M}.


The innermost dusty structure is compact, with characteristic sizes on subparsec-to-parsec scales, making it difficult to spatially resolve with current facilities for most AGNs. Long-baseline infrared interferometry has provided direct constraints on torus sizes in a limited number of nearby bright AGNs \citep{2009A&A...507L..57K,2011A&A...527A.121K,2011A&A...536A..78K,2013A&A...558A.149B,2020A&A...635A..92G,2024A&A...690A..76G}. However, such measurements remain limited by angular resolution and sensitivity. Dust reverberation mapping (RM) provides an alternative method by measuring the delayed response of infrared dust emission to variations in the dust-heating UV/optical continuum from the accretion disk \citep{1982ApJ...255..419B,1989ApJ...337..236C}. The measured lag corresponds to an effective light-travel distance, $R_{\rm dust}\simeq c\tau_{\rm dust}$, and thus provides a characteristic radius of the dust-emitting region. This RM-based radius is complementary to interferometric sizes, which trace the projected intensity distribution of the hot dust. Previous studies have found interferometric sizes to be larger than RM-based sizes by about a factor of two, likely because the two techniques weight the dust emission differently and are affected by the geometry of the innermost dust structure \citep{2011A&A...527A.121K,2013ApJ...771...87H,2014ApJ...788..159K,2020A&A...635A..92G,2023A&A...669A..14G,2024A&A...690A..76G}. Therefore, dust RM provides an efficient way to probe torus sizes in large AGN samples, especially for sources that cannot be directly spatially resolved.

Theoretically, if the inner boundary of the dusty torus is set by dust sublimation under radiative equilibrium, the incident dust-heating flux satisfies $L_{\rm heat}/(4\pi R_{\rm dust}^2)\propto T_{\rm sub}^4$. Since the dust sublimation temperature is approximately fixed, this leads to the expected scaling $R_{\rm dust}\propto L_{\rm heat}^{0.5}$ \citep{1987ApJ...320..537B,2007A&A...476..713K}. This prediction has been tested extensively with near-infrared dust RM, especially in the $K$ band, which traces hot dust close to the sublimation radius. Early $K$-band dust RM studies of nearby Seyfert~1 galaxies found a tight relation between dust radius and AGN luminosity ($R_{\rm dust}-L_{\rm AGN}$) with a slope close to the theoretical expectation of 0.5 \citep{2006ApJ...639...46S}. \citet{2014ApJ...788..159K} expanded the sample to 17 nearby Type~1 AGNs and confirmed a similar relation after correcting for accretion-disk contamination (ADC) in the near-infrared light curves. With a larger compilation of 36 Type~1 AGNs with $K$-band dust RM measurements extending to more luminous quasars, \citet{2019ApJ...886..150M} found a shallower slope of $0.42$. More recently, \citet{2024MNRAS.531.3310L} measured near-infrared dust lags for 13 AGNs at $0.3<z<0.8$ using UltraVISTA near-infrared light curves \citep{2012A&A...544A.156M}. They found that these lags are systematically below the local $R_{\rm dust}-L_{\rm AGN}$ relation but become consistent with previous results after applying an empirical rest-frame wavelength correction.

The advent of the Wide-field Infrared Survey Explorer (WISE; \citealt{2010AJ....140.1868W}) has further extended dust RM from the near-infrared $K$ band to the mid-infrared W1 ($3.4\,\mu{\rm m}$) and W2 ($4.6\,\mu{\rm m}$) bands, enabling systematic lag measurements for much larger AGN samples. \citet{2019ApJ...886...33L} demonstrated the feasibility of WISE-based dust RM using Palomar--Green quasars at $z<0.5$ and reported that the torus size correlates with AGN bolometric luminosity with slopes of $0.47$ and $0.45$ for the W1 and W2 bands, respectively. \citet{2020ApJ...900...58Y} further measured W1-band lags for 587 quasars over $0.3<z<2$, substantially increasing the number of AGNs with mid-infrared dust lag measurements. More recent dust RM studies have generally found shallower slopes of the $R_{\rm dust}-L_{\rm AGN}$ relations than the simple radiative-equilibrium expectation of 0.5. \citet{2023MNRAS.522.3439C} analyzed 78 AGNs with available H$\beta$ RM measurements and obtained $R_{\rm dust}-L_{\rm AGN}$ relations with slopes of $0.37$ in both the W1 and W2 bands. \citet{2024ApJ...968...59M} constructed a larger dust RM sample and reported best-fit slopes of $0.39$ and $0.33$ for the W1 and W2 bands, respectively. Similarly shallow $R_{\rm dust}-L_{\rm AGN}$ relations were reported by \citet{2025ApJ...993..203T} and \citet{2025ApJ...994..188A}, with W1-band (W2-band) slopes of $0.41$ ($0.40$) and $0.35$ ($0.33$), respectively.

The origin of these shallow slopes remains an open question. Several physical and observational effects may contribute to the deviation from the canonical $R_{\rm dust}\propto L_{\rm AGN}^{0.5}$ relation. These include anisotropic illumination from the accretion disk \citep{2010ApJ...724L.183K,2011ApJ...737..105K}, the delayed response of the inner dust torus to luminosity variations \citep{2009ApJ...700L.109K,2013ApJ...775L..36K}, and luminosity-dependent variations in the UV/optical continuum shape \citep{2013ApJS..206....4K}. Recent studies have further suggested that the accretion state may play an important role in regulating the $R_{\rm dust}-L_{\rm AGN}$ relation, with higher Eddington-ratio AGNs tending to exhibit smaller dust radii than expected from their luminosities \citep{2023MNRAS.522.3439C,2024ApJ...968...59M,2025ApJ...993..203T,2025ApJ...994..188A}. This trend can be explained by the self-shadowing effect of slim disks, where the geometrically thick inner disk reduces the ionizing radiation received by the dusty torus \citep{2011ApJ...737..105K,2014ApJ...797...65W,2018ApJ...856....6D}. Such accretion-rate-dependent radius offsets may contribute to the flattening of the $R_{\rm dust}-L_{\rm AGN}$ relation if the Eddington-ratio distribution varies systematically with luminosity. In addition, the choice of AGN luminosity proxy may affect the inferred $R_{\rm dust}-L_{\rm AGN}$ relation, as monochromatic luminosities or bolometric luminosities derived from simple bolometric corrections may not fully represent the dust-heating luminosity. \citet{2024A&A...690A..76G} found that the inferred $R_{\rm dust}-L_{\rm AGN}$ relation depends on the adopted AGN luminosity indicator, with different luminosity proxies yielding different slopes.

Most previous RM studies have focused on luminous AGNs hosting SMBHs, leaving the low-mass and low-luminosity regime poorly explored. Intermediate-mass black holes (IMBHs) and low-mass black holes provide important links between stellar-mass black holes and SMBHs and may retain valuable information about black hole seeding and the early growth of SMBHs \citep{2017IJMPD..2630021M,2020ARA&A..58..257G,2020ARA&A..58...27I,2022NatAs...6...26R}. However, identifying and characterizing IMBHs and low-mass black hole systems remains challenging because their spheres of influence are generally too small to be spatially resolved, while single-epoch virial mass estimates based on broad Balmer lines are subject to substantial uncertainties and possible contamination from transient stellar processes \citep{2004ApJ...610..722G,2007ApJ...670...92G,2012ApJ...755..167D,2018ApJ...863....1C}. Their faint nuclear emission, strong host-galaxy dilution, and short characteristic variability and reverberation timescales further complicate systematic RM studies \citep{2024ApJ...974..288Z,2024ApJ...976..116P}.

\begin{figure*}[t!]
    \centering
    \includegraphics[width=0.95\textwidth]{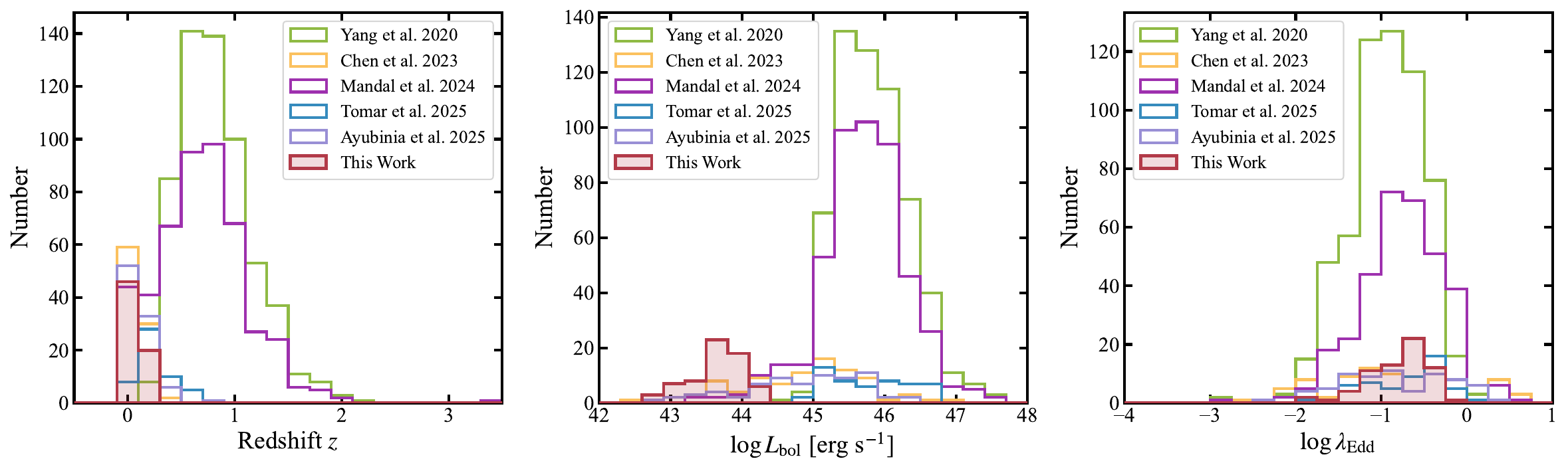} 
    \caption{Distributions of redshift (left), bolometric luminosity (middle), and Eddington ratio (right) for our core sample and previous dust RM samples. The sample analyzed in this work is shown as the filled histogram with burgundy color, while samples compiled from previous studies are shown as colored open histograms.}
    \label{fig:distribution}
\end{figure*}

Both BLR and continuum RM have been successfully applied to the prototypical IMBH NGC~4395. BLR RM measurements based on C\,{\sc iv} and H$\alpha$ have revealed minute-to-hour-scale lags and yielded black hole mass estimates of approximately $10^{4}$--$10^{5}\,M_\odot$ \citep{2005ApJ...632..799P,2019NatAs...3..755W,2020ApJ...892...93C,2021ApJ...921...98C,2024MNRAS.530.3578G,2024ApJ...976..116P}. Continuum RM studies have also detected wavelength-dependent interband lags and constrained the size and structure of the compact accretion disk in this source \citep{2006ApJ...650...88D,2022ApJ...934L..37M,2023MNRAS.519.3366M,2026arXiv260114787P}. Previous near-infrared dust RM studies have also measured K-band lags in several nearby low-luminosity Seyfert galaxies, including three low-mass AGNs (NGC~4051, NGC~4593, and NGC~3227) with reported black hole masses below \(10^7\,M_\odot\) \citep{2006ApJ...639...46S,2014ApJ...788..159K}. Nevertheless, dust RM measurements remain sparse in the low-mass and low-luminosity regime. More recently, \citet{2025ApJ...989L..26S} reported the first detections of mid-infrared dust lags in the prototypical IMBHs NGC~4395 and POX~52. Their measurements broadly follow the extrapolation of the $R_{\rm dust}-L_{\rm AGN}$ relation established for more luminous AGNs hosting SMBHs, demonstrating the feasibility of dust RM in this regime.

Recently, \citet{2026HXG} compiled a literature-based parent catalog of 1447 IMBH candidates with reported black hole mass estimates of
$\log(M_{\rm BH}/M_\odot) \lesssim 6.5$. Based on this parent catalog, we perform a systematic dust RM analysis and select a core sample of 65 sources based on the quality of their optical and infrared light curves and lag measurements. The core sample is dominated by AGNs with black hole masses above $\sim10^6\,M_\odot$. For lower-mass AGNs, the fainter nuclear emission and stronger host-galaxy dilution generally make lag measurements more difficult and require more careful treatment of the light curves. Liu et al. (in prep.) will present a more detailed variability and dust-lag analysis specifically targeting sources with black hole masses below $10^6\,M_\odot$. In this work, we focus on the statistical properties of the dust-lag measurements and investigate the $R_{\rm dust}-L_{\rm AGN}$ relation in the low-mass and low-luminosity regime. The structure of this paper is as follows. In Section~\ref{sec:data}, we describe the low-mass AGN sample, the optical and infrared data, and the construction of the light curves. In Section~\ref{sec:Analysis and Results}, we measure the dust lags using three independent methods. We further assess the reliability of the lag measurements and apply corrections for redshift and possible contamination from the accretion disk. In Section~\ref{sec:discussion}, we compare the lag measurements obtained from different methods and the torus sizes in the W1 and W2 bands. We also investigate the $R_{\rm dust}-L_{\rm AGN}$ relation for our low-mass AGN sample, perform a joint analysis with previous dust RM samples, and discuss the dust RM in a tidal disruption event (TDE) candidate. Finally, we summarize our main conclusions in Section \ref{sec:Summary}.

\begin{table*}[t]
    \begin{center}
    \caption{Properties of the Low-Mass AGNs in the Core Sample.}
    \vspace{-1em}
    \label{tab:properties}
    \setlength\tabcolsep{2.4mm}
    \begin{tabular}{cccccccccc}
        \toprule
        ID  &Name& RA & DEC & $z$ & $\log L_{5100}$ & $\log L_{5100,\mathrm{AGN}}$ & $\log L_{\mathrm{bol}}$ & $\log M_{\mathrm{BH}}$ & $\log \lambda_{\mathrm{Edd}}$ \\
         && (deg) & (deg) &  & (erg s$^{-1}$) & (erg s$^{-1}$) & (erg s$^{-1}$) & ($M_\odot$) &  \\
 (1) & (2)& (3)& (4)& (5)& (6)& (7)& (8)& (9)&(10)\\
        \midrule
        1  &J003221.03+242359.3
& 8.0876& 24.3998
& 0.0656& 43.17& 42.63& 43.82& 6.35& $-$0.6304
\\
        2  &J003911.37+271806.1
& 9.7974& 27.3017
& 0.1419& 43.26& 42.67& 43.78& 6.11& $-$0.4339
\\
        3  &J010712.03+140844.9& 16.8001& 14.1458& 0.0767& 43.03& 42.49& 43.65& 6.20& $-$0.6452\\
        \bottomrule
    \end{tabular}
    \end{center}
\textbf{Note.} Column (1): Object ID number. Column (2): Object name. Column (3): Right ascension. Column (4): Declination. Column (5): Redshift. Columns (6)--(8): Logarithmic luminosity measurements, including the total monochromatic luminosity at 5100\,\AA, the host-subtracted AGN luminosity at 5100\,\AA, and the AGN bolometric luminosity. Column (9): Logarithm of the black hole mass in units of $M_\odot$. Column (10): Logarithm of the Eddington ratio. The values in Columns (3)–(9) are taken from \citet{2026HXG}. The Eddington ratio in Column (10) is calculated from the AGN bolometric luminosity and the black hole mass. This table is available in its entirety in machine-readable form in the online article.
\end{table*}

\section{Sample and Data} \label{sec:data}
\subsection{Low-Mass AGN Sample}\label{subsec:IMBH sample}
In this work, the parent sample is drawn from the IMBH candidate catalog compiled by \citet{2026HXG}, who collected 1447 IMBH candidates with previously reported black hole masses of $\log(M_{\rm BH}/M_\odot)\leq6.5$ from the literature \citep[e.g.,][]{2004ApJ...610..722G,2007ApJ...670...92G,2012ApJ...755..167D,2018ApJS..235...40L,2018ApJ...863....1C,2025arXiv250312898L} and uniformly reanalyzed the available SDSS spectra using consistent spectral decomposition and virial mass calibration procedures. From this catalog, we select 65 sources as the core sample based on a series of selection criteria. These criteria mainly involve the quality of the optical and infrared light curves and lag measurements. The details of the selection procedure and quality assessment are described in Section~\ref{subsec:Quality Assessment}. 

The basic properties of the sources in the core sample are listed in Table~\ref{tab:properties}. Figure~\ref{fig:distribution} presents a comparison between the redshift, AGN bolometric luminosity, and Eddington ratio distributions of our core sample and those reported in previous dust RM studies \citep{2020ApJ...900...58Y,2023MNRAS.522.3439C,2024ApJ...968...59M,2025ApJ...993..203T,2025ApJ...994..188A}. For our sample, we adopt the black hole masses from the uniform reanalysis of \citet{2026HXG}, in which $M_{\rm BH}$ is estimated from the luminosity and width of the broad H$\alpha$ component. The bolometric luminosity is derived from the broad H$\alpha$ luminosity following the empirical relation of \citet{2005ApJ...630..122G}, $L_{\rm H\alpha}=5.25\times10^{42}(L_{5100}/10^{44})^{1.157}$, and the bolometric correction $L_{\rm bol}=9.26 \times L_{5100}$ \citep{2006ApJS..166..470R}. The Eddington ratio is defined as $\lambda_{\rm Edd}=L_{\rm bol}/L_{\rm Edd}$, where $L_{\rm Edd}=1.26\times10^{38}(M_{\rm BH}/M_\odot)$ erg s$^{-1}$. Our core sample extends dust RM measurements toward lower black hole masses and luminosities, with $\log(M_{\rm BH}/M_\odot)\sim4.5-7.0$, and $\log(L_{\rm bol}/{\rm erg\,s^{-1}})\sim41.1-44.4$. Compared with previous dust RM samples, our sources cover a broadly overlapping range of Eddington ratios, although high-Eddington-ratio objects with $\log\lambda_{\rm Edd}>-0.3$ are less common. This sample provides an opportunity to test whether the $R_{\rm dust}-L_{\rm AGN}$ relation established for more luminous AGNs can be extended into the low-mass and low-luminosity regime, while also allowing us to explore the possible dependence of dust structures on accretion state.

\subsection{Optical Data}\label{subsec:Optical Data}
To obtain optical light curves for the sources in our parent sample, we consider various transient surveys, including the All-Sky Automated Survey for Supernovae (ASAS-SN; \citealt{2014AAS...22323603S}), the Catalina Real-Time Transient Survey (CRTS; \citealt{2009ApJ...696..870D}), the Palomar Transient Factory (PTF; \citealt{2009PASP..121.1395L}), and the Zwicky Transient Facility (ZTF; \citealt{2019PASP..131a8002B}). These surveys provide long-term optical monitoring with different filters, cadences, and limiting magnitudes.

Previous dust RM studies typically combine optical data from multiple surveys to extend the temporal baseline, which is important for luminous AGNs with dust lags ranging from hundreds to thousands of days. In contrast, the sources in our sample are less luminous, and their expected dust lags are much shorter, typically ranging from a few days to several hundred days. In addition, ASAS-SN has a relatively shallow limiting magnitude ($V\sim17$ mag; $g\sim17.5$ mag), while the cadence of CRTS and PTF is relatively sparse ($\sim5$--20 days). As a result, we find that the light curves from these surveys are generally insufficient in quality for reliable lag measurements for most sources in our sample.

Therefore, we use optical light curves from the ZTF, which provides observations in the $g$, $r$, and $i$ bands with a typical cadence of $\sim3$ days and median depths of $g\sim20.8$, $r\sim20.6$, and $i\sim19.9$ mag (5$\sigma$ in 30~s). For our analysis, we incorporate the ZTF $g$- and $r$-band light curves from Data Release 23, covering the period from 2018 to 2024. To ensure data quality, we select measurements with \texttt{catflags}=0 to exclude data affected by clouds, moonlight, or poor seeing conditions. We further apply a $3\sigma$ clipping procedure to remove outliers and bin the light curves into one-day intervals to improve the signal-to-noise ratio. All magnitudes are converted to flux using the photometric zero points provided by the SVO Filter Profile Service \footnote{\url{https://svo2.cab.inta-csic.es/svo/theory/fps3/index.php?mode=browse&gname=Palomar&gname2=ZTF&asttype=}}.

\subsection{Infrared Data}\label{subsec:Infrared Data}
Infrared data are obtained from the WISE, a space-based infrared telescope that performed an all-sky survey in four bands at 3.4, 4.6, 12, and 22 $\mu$m (W1--W4). After the completion of the primary mission, the telescope was reactivated in 2013 as the Near-Earth Object Wide-field Infrared Survey Explorer (NEOWISE; \citealt{2011ApJ...743..156M}) mission, which continued scanning the sky in the W1 (3.4 $\mu$m) and W2 (4.6 $\mu$m) bands. The WISE/NEOWISE survey observes the entire sky approximately every six months, providing long-term infrared monitoring suitable for dust RM studies.

In this work, we retrieve single-exposure profile-fit photometry in the W1 and W2 bands from the WISE and NEOWISE databases via the NASA/IPAC Infrared Science Archive. To ensure reliable photometry, we select measurements satisfying the following quality criteria: best frame image quality (\texttt{qi\_fact=1}), sufficient separation from the South Atlantic Anomaly (\texttt{saa\_sep$\ge$5}), no moon contamination (\texttt{moon\_masked=0}), and no spurious detections (\texttt{cc\_flags=0}). The WISE magnitudes are converted to flux using the zero-point flux densities from \citet{2013AJ....145....6J}, and the W1 and W2 measurements are binned into 180-day intervals to construct evenly sampled infrared light curves. To match the time coverage of the ZTF optical light curves, we only retain WISE data obtained with $\mathrm{MJD} > 58000$ in the subsequent analysis.

\subsection{Light Curves}\label{subsec:Light Curves}
To construct reliable optical and infrared light curves for lag measurements, we impose requirements on the temporal sampling of the light curves. We select sources with more than 100 epochs in the ZTF $g$- and $r$-band light curves (after one-day binning) and more than 10 epochs in the WISE W1- and W2-band light curves (after 180-day binning). This selection yields 1033 sources from the parent sample. 

The optical light curves are constructed from ZTF $g$- and $r$-band data. To correct for systematic offsets between different bands and observing conditions, we perform intercalibration using the Python-based Bayesian package \texttt{PyCALI} \citep{2014ApJ...786L...6L}\footnote{\url{https://github.com/LiyrAstroph/PyCALI}}. \texttt{PyCALI} models AGN variability with a damped random walk (DRW; \citealt{2009ApJ...698..895K,2010ApJ...721.1014M}) process and determines the optimal multiplicative and additive scaling factors between light curves. In this process, we adopt the ZTF $g$-band light curve as the reference, because it typically has better sampling for most sources; otherwise, the ZTF $r$-band is used. The interband continuum lags between the $g$ and $r$ bands are much shorter than the infrared lag timescales and are therefore neglected in the intercalibration. We further identify and remove outliers based on the DRW reconstruction, excluding data points that deviate by more than $5\sigma$ from the reconstructed light curves. To ensure consistent sampling and improve the signal-to-noise ratio, the calibrated optical light curves are binned using a 10-day window. For NGC 4395, we adopt a shorter 3-day binning window, consistent with \citet{2025ApJ...989L..26S}, due to its much shorter expected dust lag. The final calibrated and binned light curves are used for subsequent lag measurements.

\section{Data Analysis}\label{sec:Analysis and Results}
In this section, we analyze the optical and infrared light curves for 1033 sources. We measure the dust lags using three independent methods: the Interpolated Cross-Correlation Function (ICCF; \citealt{1998PASP..110..660P}), JAVELIN \citep{2011ApJ...735...80Z}, and PYROA \citep{2021MNRAS.508.5449D}. The reliability of the lag measurements is then evaluated using a series of quality criteria, and a core sample of 65 sources is selected for subsequent analysis. Finally, we apply corrections for redshift and ADC in the infrared light curves to obtain the corrected lag measurements.

\subsection{Lag Measurement}\label{subsec:Dust Lag Measurements}

\begin{figure*}[t]
    \centering
    \includegraphics[width=0.95\textwidth]{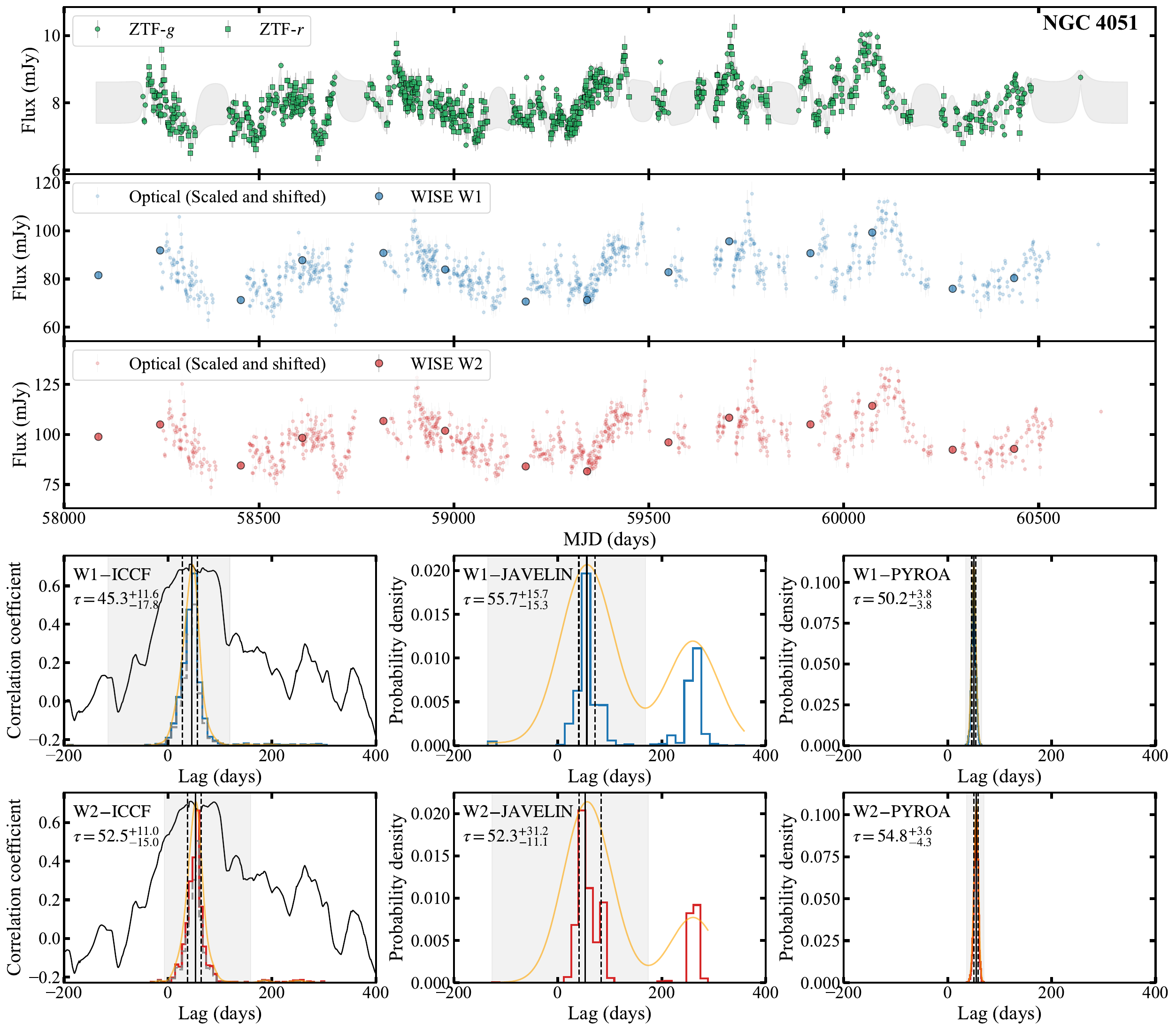} 
    \caption{Light curves and lag measurements for NGC 4051. The top three panels show the optical and infrared light curves. The first panel presents the intercalibrated optical light curves, with the gray shaded region representing the DRW model reconstructed by PyCALI. The second and third panels show the accretion-disk-corrected infrared light curves in the WISE W1 and W2 bands, respectively. The blue and red transparent points represent the scaled optical light curves shifted by the ICCF centroid lag relative to the W1 and W2 light curves, respectively. The fourth row shows the W1-band lag measurements obtained using different methods. The left panel presents the CCF (black curve), together with the unweighted CCCD (blue solid histogram) and the weighted CCCD (gray dashed histogram) derived from ICCF. The orange curve shows the smoothed weighted lag distribution, and the gray shaded region indicates the range of the primary peak. The middle and right panels show the lag distributions obtained from JAVELIN and PYROA, respectively. The fifth row shows the W2-band lag measurements derived from ICCF (left), JAVELIN (middle), and PYROA (right) in the same format as the fourth row.}
    \label{fig:result_35}
\end{figure*}

To measure the dust lags between the optical and infrared light curves, we employ three independent methods: ICCF, JAVELIN, and PYROA. In this work, the ICCF results are adopted as the primary lag measurements, while the JAVELIN and PYROA results are used as independent consistency checks.

The ICCF measurements are performed using a Python-based improved interpolated cross-correlation function package\footnote{\url{https://github.com/legolason/PyIICCF}} (\texttt{PyI$^2$CCF}; see \citealt{2021AAS...23722608G,2022ApJ...925...52U}, for details). The ICCF method estimates the correlation between two unevenly sampled light curves by linearly interpolating one light curve and cross-correlating it with the other over a range of trial time delays. Following common practice in RM studies, we adopt the centroid lag, $\tau_{\rm cent}$, computed from all correlation coefficients exceeding 80\% of the peak correlation coefficient, as the lag measurement \citep{2004ApJ...613..682P}. The lag uncertainties are estimated using the flux randomization and random subset selection (FR/RSS) methods \citep{1998PASP..110..660P}. For each source, 10,000 FR/RSS realizations are generated to construct the cross-correlation centroid distribution (CCCD). The median of the CCCD is adopted as the final ICCF lag, while the 16th and 84th percentiles are taken as the corresponding $1\sigma$ uncertainties.

In our implementation, we adopt the standard ICCF two-way interpolation scheme, where each light curve is interpolated in turn and the resulting correlation functions are combined. Given the substantially different sampling densities between the optical and infrared light curves, we further tested a one-directional ICCF calculation by retaining only the interpolation of the sparsely sampled infrared light curve onto the densely sampled optical epochs. In addition, we examined the effect of replacing the FR/RSS method with an FR-only approach, since RSS may reduce the number of infrared epochs in each realization. These tests show that the lag measurements remain broadly consistent for most sources, while some sources exhibit smaller lag uncertainties under the tested configurations.

We make use of the weighting scheme implemented in \texttt{PyI$^2$CCF} to mitigate the influence of seasonal gaps and alias peaks in the lag distribution. Similar to previous studies \citep{2017ApJ...851...21G,2019ApJ...887...38G}, the weight function consists of two components: (1) the autocorrelation function of the optical light curve, which characterizes the intrinsic variability timescale of the source, and (2) an overlapping function that quantifies the fraction of overlapping data points between the two light curves at a given trial lag. The final weight is obtained by convolving these two components and is applied to the CCCD. This weighting scheme effectively suppresses lag solutions that are strongly affected by seasonal sampling patterns and helps identify the preferred lag peak \citep{2024ApJS..272...26S,2024ApJS..275...13W}.

We also employ a parallel technique, JAVELIN\footnote{\url{https://github.com/nye17/javelin}} \citep{2011ApJ...735...80Z}, to estimate the dust lags for comparison. JAVELIN models the optical light curve as a damped random walk (DRW) process, which has been widely used to describe AGN variability \citep{2009ApJ...698..895K,2010ApJ...721.1014M}, although deviations have been reported on very short (minutes) and long (decades) timescales \citep{2011ApJ...743L..12M,2017ApJ...847..132G,2024ApJ...969...78S}. The infrared light curve is modeled as a scaled, smoothed, and time-delayed response to the optical continuum through a top-hat transfer function characterized by a lag and width. The model parameters are constrained within a Bayesian framework using Markov Chain Monte Carlo (MCMC) simulations, which provide posterior probability distributions for the lag measurements. For each JAVELIN run, we adopt $n_{\rm chain}=n_{\rm walkers}=n_{\rm burn}=100$, corresponding to a total of 10,000 MCMC realizations, to ensure adequate sampling of the posterior distribution. 

As another independent consistency check of the ICCF lag measurements, we also measure the dust lags using the PYROA\footnote{\url{https://github.com/FergusDonnan/PyROA}} \citep{2021MNRAS.508.5449D}. PYROA models AGN variability using the running optimal average (ROA) method, which represents an inverse-variance weighted average of all data points. The optical and infrared light curves are jointly fitted by shifting and scaling the infrared light curve relative to the optical light curve. The ROA model is multiplied by a Gaussian window function with width $\Delta$, which suppresses the influence of distant data points and controls the effective model complexity. Unlike JAVELIN, PYROA does not rely on the commonly adopted DRW assumption for AGN variability, making it a useful independent test of the lag measurements. The model parameters are determined within a Bayesian framework using MCMC simulations. For each source, we generate 10,000 MCMC realizations and adopt the ICCF lag as the initial lag guess to assist convergence.

\begin{figure*}[t]
    \centering
    \includegraphics[width=0.95\textwidth]{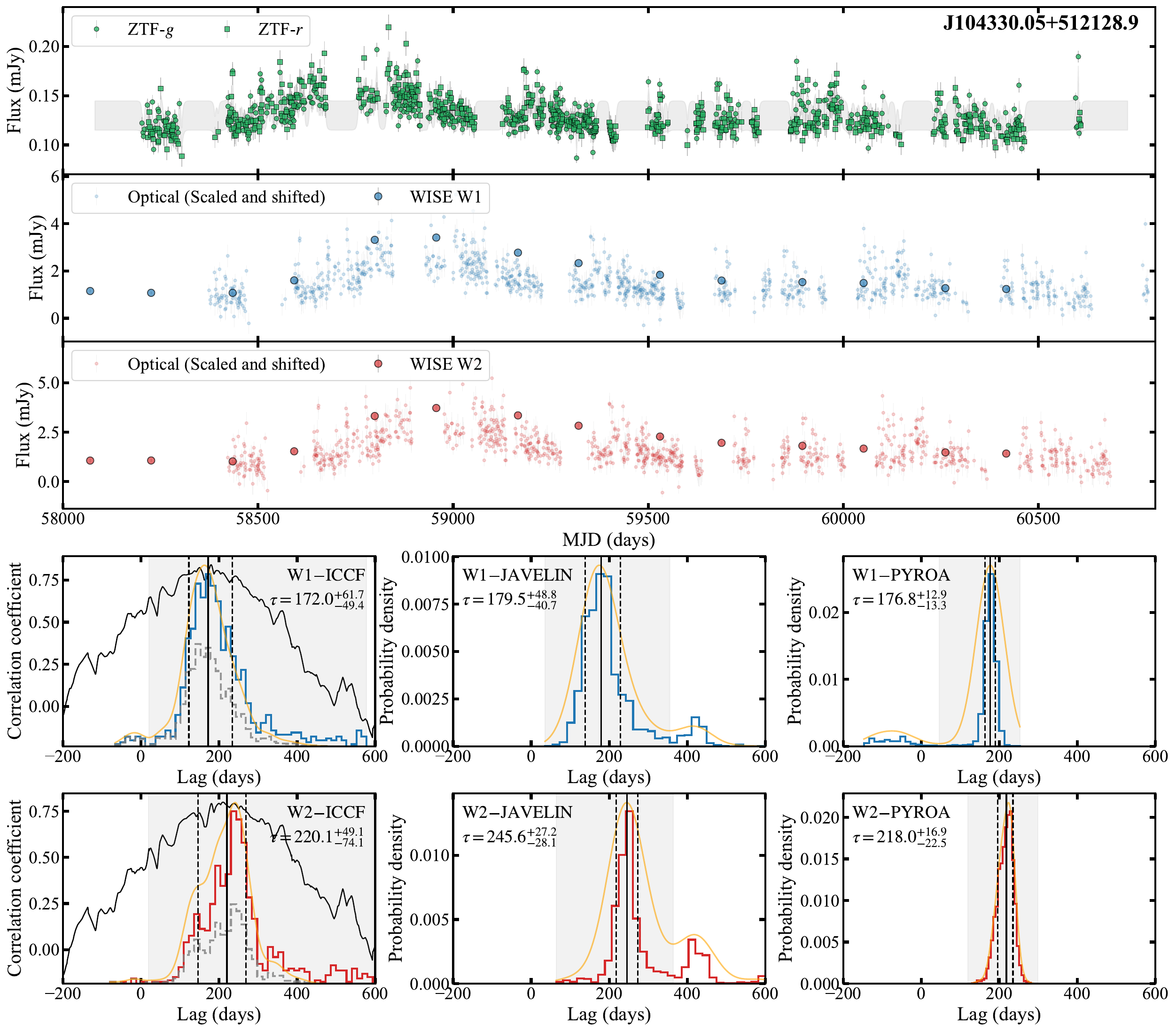} 
    \caption{Same as Figure \ref{fig:result_35}, but for J104330.05+512128.9, a source associated with a TDE-like transient.}
    \label{fig:result_25}
\end{figure*}

For all three lag-measurement methods, we adopt the same lag-searching window to facilitate a direct comparison among the results. In principle, a wider lag-searching window allows a broader exploration of possible lag solutions, while in practice, it tends to introduce additional alias peaks and may even hinder the recovery of the true lag \citep{2024ApJS..275...13W}. To determine an appropriate and uniform lag-searching window for our sample, we estimate the expected dust lags for each source using the bolometric luminosities provided by \citet{2026HXG} together with the $R_{\rm dust}-L_{\rm AGN}$ relations from \citet{2024ApJ...968...59M}. The expected W1-band lags span from 1.3 to 154.6 days, with 90\% of the sample falling between 14.2 and 103.5 days. The expected W2-band lags range from 4.3 to 226.4 days, with 90\% of the sample distributed between 30.5 and 162.2 days. Based on these expectations, we adopt a uniform lag-searching window of $[-200,400]$ days for all sources and all three methods, which comfortably encompasses the expected lag range of the sample. For the small number of sources with measured lags approaching or exceeding 200 days, we repeat the lag measurements using an extended lag-searching window of $[-200,600]$ days. An exception is NGC 4395, whose dust lags have previously been measured by \citet{2025ApJ...989L..26S} using a lag-searching window of $\pm15$ days. For consistency with the literature and to enable a direct comparison of the lag measurements, we adopt the same lag-searching window for this source.

We note that all three methods can exhibit multiple peaks in their lag posterior distributions owing to limited temporal sampling, seasonal gaps, quasi-periodic variability, etc. \citep{2017ApJ...851...21G,2019ApJ...887...38G,2020ApJ...901...55H,2024ApJS..275...13W,2025ApJ...994..246W}. To identify the most probable lag solution, we smooth the lag distributions using a Gaussian kernel density estimator (KDE). The initial KDE bandwidth is determined using Scott's rule \citep{Scott1992}, which sets the smoothing scale based on the dispersion and effective number of lag samples. In a small number of cases, modest adjustments to the bandwidth are applied to suppress numerical fluctuations and facilitate the identification of the primary lag peak. The highest peak in the smoothed distribution is identified as the primary lag peak. We then locate the nearest local minima on both sides of this peak and adopt them as the boundaries of the primary lag region. Only lag samples within this region are retained for the final lag determination, while samples associated with secondary peaks are discarded. The final lag and its uncertainties are then calculated from the remaining samples using the 16th, 50th, and 84th percentiles of the corresponding lag distribution.

\begin{sidewaystable}
\begin{center}
\caption{Lag Measurements for the W1 and W2 Bands Derived from Different Methods.}
\vspace{-1em}
\renewcommand{\arraystretch}{1.0}
\label{tab:lag}
\setlength\tabcolsep{1.6mm}{
\begin{tabular}{ccccccccccccc}
\toprule
\toprule
ID & Name & Case & $\tau_{\mathrm{W1,ICCF}}$ & $R_{\mathrm{max,W1}}$ & $p_{\mathrm{W1}}$ & $\tau_{\mathrm{W1,JAVELIN}}$ & $\tau_{\mathrm{W1,PYROA}}$ & $\tau_{\mathrm{W2,ICCF}}$ & $R_{\mathrm{max,W2}}$ & $p_{\mathrm{W2}}$ & $\tau_{\mathrm{W2,JAVELIN}}$ & $\tau_{\mathrm{W2,PYROA}}$ \\
(1) & (2) & (3) & (4) & (5) & (6) & (7) & (8) & (9) & (10) & (11) & (12) & (13)\\
\hline

\multirow{2}{*}{1} & \multirow{2}{*}{J003221.03+242359.3} 
& Non-ADC 
& $20.62_{-23.2}^{+82.6}$ & 0.82 & 0.000 
& $17.90_{-15.4}^{+12.3}$ & $23.51_{-13.4}^{+11.3}$ 
& $68.71_{-64.9}^{+44.7}$ & 0.74 & 0.068 
& $39.19_{-36.1}^{+24.9}$ & $44.33_{-5.9}^{+4.5}$ \\

& & ADC 
& $27.96_{-28.9}^{+90.6}$ & 0.80 & 0.016 
& $34.18_{-15.1}^{+12.1}$ & $27.17_{-18.8}^{+11.7}$ 
& $75.81_{-57.5}^{+43.6}$ & 0.71 & 0.094 
& $42.74_{-20.2}^{+21.6}$ & $55.67_{-16.7}^{+22.1}$ \\

\midrule

\multirow{2}{*}{2} & \multirow{2}{*}{J003911.37+271806.1} 
& Non-ADC 
& $98.81_{-47.7}^{+67.9}$ & 0.71 & 0.030 
& $98.03_{-50.0}^{+73.9}$ & $84.06_{-130.8}^{+110.9}$ 
& $96.92_{-56.7}^{+81.5}$ & 0.70 & 0.020 
& $101.96_{-79.2}^{+110.3}$ & $78.15_{-42.5}^{+9.8}$ \\

& & ADC 
& $100.57_{-42.2}^{+78.3}$ & 0.72 & 0.025 
& $107.90_{-46.9}^{+63.0}$ & $79.52_{-18.8}^{+16.4}$ 
& $98.32_{-68.6}^{+75.9}$ & 0.70 & 0.021 
& $96.99_{-77.2}^{+104.7}$ & $83.84_{-34.7}^{+32.4}$ \\

\midrule

\multirow{2}{*}{3} & \multirow{2}{*}{J010712.03+140844.9} 
& Non-ADC 
& $86.72_{-65.3}^{+59.3}$ & 0.74 & 0.093 
& $71.46_{-38.6}^{+35.5}$ & $87.60_{-52.3}^{+48.1}$ 
& $189.84_{-151.8}^{+105.3}$ & 0.79 & 0.092 
& $186.35_{-125.9}^{+151.3}$ & $198.50_{-109.1}^{+93.5}$ \\

& & ADC 
& $98.05_{-58.2}^{+111.2}$ & 0.74 & 0.094 
& $83.88_{-33.1}^{+35.6}$ & $109.42_{-33.6}^{+41.4}$ 
& $193.87_{-158.0}^{+101.3}$ & 0.79 & 0.062 
& $203.30_{-140.9}^{+178.2}$ & $199.19_{-90.7}^{+97.6}$ \\

\bottomrule
\end{tabular}}
\end{center}

\vspace{-0.5em}
\textbf{Note.} Column (1): Object ID number. Column (2): Object name. Column (3): ADC case. Columns (4), (7), and (8): W1-band lag measurements obtained from ICCF, JAVELIN, and PYROA, respectively. Columns (5) and (6): Maximum cross-correlation coefficient and probability value indicating the robustness of ICCF lag measurements in the W1 band. Columns (9), (12), and (13): W2-band lag measurements obtained from ICCF, JAVELIN, and PYROA, respectively. Columns (10) and (11): Maximum cross-correlation coefficient and probability value indicating the robustness of ICCF lag measurements in the W2 band. This table is available in its entirety in machine-readable form in the online article.
\end{sidewaystable}

As illustrative examples of the lag measurements in our core sample, Figures~\ref{fig:result_35} and \ref{fig:result_25} show the results for NGC~4051 and J104330.05+512128.9, respectively. NGC~4051 is a nearby low-mass AGN with BLR, UV/optical continuum, and near-infrared dust RM measurements \citep{2006ApJ...639...46S,2009ApJ...702.1353D,2017ApJ...840...97F,2026MNRAS.548ag642M}. It exhibits relatively short W1- and W2-band lags in our sample. J104330.05+512128.9 is associated with the nuclear transient AT2019dqv, which has been discussed as a possible TDE-like accretion flare \citep{2024MNRAS.529.2559V,2025A&A...695A.228N}. This source exhibits relatively long W1- and W2-band lags in our sample, and its dust reverberation properties are further discussed in Section~\ref{subsec:TDE}. For both sources, the W1- and W2-band lag measurements obtained from the ICCF, JAVELIN, and PYROA methods are broadly consistent within the uncertainties. After shifting the optical light curves by the corresponding ICCF centroid lags, the optical variations show good agreement with the W1 and W2 light curves, and the corresponding $R_{\rm max}$ values are close to or exceed 0.8. The lag measurements for all sources in the core sample are summarized in Table~\ref{tab:lag}. 

\begin{table*}[t]
    \begin{center}
    \caption{Quality Selection of Lag Measurements.}
    \vspace{-1em}
    \label{tab:Selection}
    \setlength\tabcolsep{8mm}{
    \begin{tabular}{lc}
        \toprule
        \toprule
        Criteria& Number\\
        \midrule
        Parent Sample: IMBH candidate catalog from \citet{2026HXG}& 1447\\
        
        Good light curves: ZTF g/r (1d-bin) epochs $>$ 100 $\&$ WISE W1/W2 (180d-bin) epochs $>$ 10 & 1033\\
        
        Strong correlation: $R_{\max,\rm W1}>0.6  \ \& \ R_{\max, \rm W2}>0.6 $ & 158\\
        High significance: $p_{\rm W1}< 0.1 \ \& \ p_{\rm W2}< 0.1 $&92\\
        Positive lags from all three methods in both the W1 and W2 bands: $\tau_{\rm W1}> 0 \ \& \ \tau_{\rm W2} > 0  $ & 65\\
        \bottomrule
    \end{tabular}}
    \end{center}
\end{table*}

In the Appendix, we present six additional representative examples, shown in Figure~\ref{fig:Lag_results}, spanning a broad range of bolometric luminosities and black hole masses in the core sample. To further assess the robustness of the lag measurements against the optical sampling strategy, we examine the impact of different optical binning schemes. In Figure~\ref{fig:lag_bin_comparison}, we present two representative sources, exhibiting relatively small and large dust lags, respectively. We compare their W1- and W2-band lag measurements derived from optical light curves binned using 3-day, 5-day, 10-day, and 15-day intervals. Overall, the lag measurements, particularly those obtained from the ICCF method, remain broadly consistent across different optical binning schemes. The 10-day and 15-day binning schemes generally produce smoother cross-correlation functions (CCFs) with more stable primary peaks. In contrast, the 3-day and 5-day binning schemes tend to yield noisier CCFs, while the corresponding lag distributions occasionally exhibit secondary peaks or pronounced asymmetries. We also visually inspect additional sources in the sample and find similar behavior in most cases. Previous dust RM studies have adopted 15-day binning for optical light curves \citep{2023MNRAS.522.3439C,2025ApJ...993..203T}. While longer binning intervals can help suppress photometric noise, they inevitably reduce the temporal resolution. Given that some sources in our sample may exhibit dust lags as short as $\sim$15 days, we therefore adopt the 10-day binning scheme as a compromise between suppressing noise-induced features and preserving sufficient temporal resolution for reliable lag measurements.

\subsection{Quality Assessment}\label{subsec:Quality Assessment}
The sparse sampling and seasonal gaps in the infrared light curves, coupled with the limited number of data points in the optical-IR light curve pairs, can produce aliasing and spurious peaks in the lag distributions. A robust quality assessment is therefore essential before adopting these lag measurements for subsequent analysis. Following previous dust RM studies \citep{2024ApJ...968...59M, 2025ApJ...993..203T, 2025ApJ...994..188A}, we applied a series of quality-selection criteria to the 1033 sources with sufficiently sampled optical and infrared light curves to select reliable dust lags, as summarized in Table~\ref{tab:Selection}.

As the first quality criterion, we required a strong correlation between the optical and infrared light curves, adopting a maximum correlation coefficient of $R_{\rm max}>0.6$ in both the W1 and W2 bands. This criterion excluded the majority of sources with weak optical–IR correlations and retained 158 candidates for further quality assessment. Although $R_{\rm max}$ quantifies the correlation strength between the optical and infrared light curves, it alone cannot guarantee the reliability of the measured lags. A high $R_{\rm max}$ may arise from the chance alignment of a small number of features, particularly in sparsely sampled or smoothly varying light curves.

To further evaluate the statistical significance of the lag measurements, we employed the null-hypothesis test implemented in the \texttt{PyI$^2$CCF} package. For each of the 158 sources passing the $R_{\rm max}>0.6$ selection, we generated 1,000 pairs of independent mock optical and infrared light curves based on the DRW model, while preserving the same sampling pattern, cadence, and photometric uncertainties as the observed light curves. The mock light curves were then analyzed using the same ICCF procedure and lag-search window as adopted for the observed data. We quantified the lag significance using the probability $p$, defined as the fraction of simulations in which a positive lag ($\tau>0$) is obtained together with a maximum correlation coefficient exceeding the observed $R_{\rm max}$. This quantity represents the probability that a positive lag with comparable correlation strength arises purely from unrelated stochastic variability. Previous dust RM studies have commonly adopted $p<0.2$ \citep{2024ApJ...968...59M,2025ApJ...993..203T}, while \citet{2025ApJ...994..188A} adopted a stricter threshold of $p\leq0.1$. Following the latter, we required $p<0.1$ in both the W1 and W2 bands. This criterion further reduced the sample from 158 to 92 sources.

As a final quality criterion, we required all three methods (ICCF, JAVELIN, and PYROA) to yield positive lags in both the W1 and W2 bands. Physically, the infrared emission from the dust torus is expected to reverberate in response to variations of the optical continuum with a positive lag. Statistically, requiring all three independent lag determination methods to yield positive lags reduces the likelihood that the measured lags are driven by the limitations of any individual method. Applying this criterion reduced the sample from 92 to 65 sources, which constitutes the core sample for the subsequent analyses.

\begin{figure*}[t]
    \centering
    \includegraphics[width=0.90\textwidth]{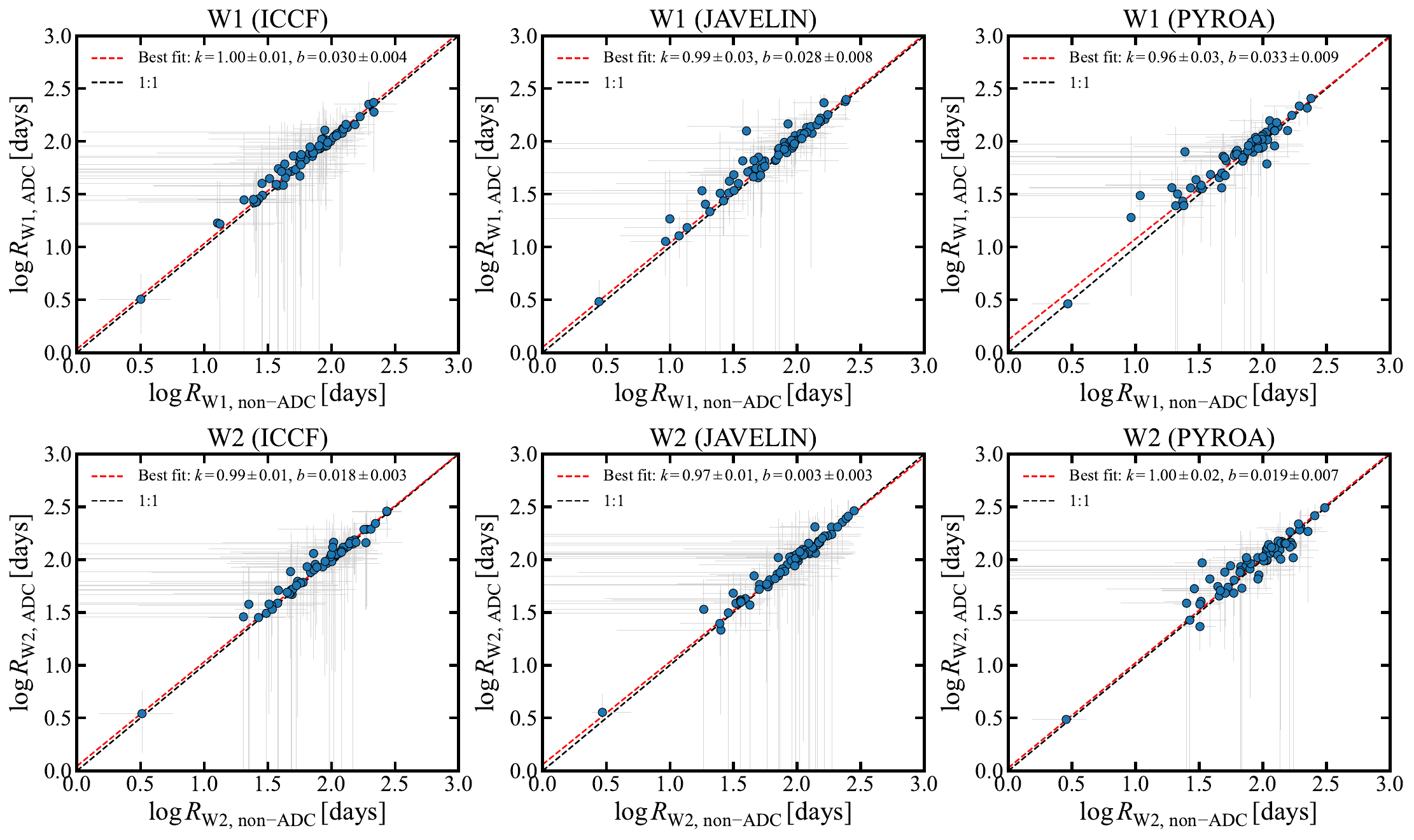} 
    \caption{Comparison of dust lags with and without ADC correction. The top and bottom rows show the results for the W1 and W2 bands, respectively. The left, middle, and right columns correspond to the lags measured using ICCF, JAVELIN, and PYROA, respectively. The red dashed line shows the best-fit relation, while the black dashed line indicates the 1:1 relation.}
    \label{fig:Lag_comparison_ADC}
\end{figure*}

\subsection{Redshift Correction}\label{subsec:Redshift Correction}
The observed dust lags are affected by both cosmological time dilation and the wavelength-dependent dust structure. At a fixed observed band, higher-redshift sources are probed at shorter rest-frame wavelengths, which trace hotter dust located closer to the central engine and therefore correspond to shorter intrinsic dust lags \citep{2001ASPC..224..149O,2014ApJ...784L..11Y,2019ApJ...886..150M,2020AJ....159..259S,2023MNRAS.522.3439C,2024ApJ...968...59M}.

In this work, we correct the observed-frame lags to rest-frame lags using $\tau_{\rm rest}=\tau_{\rm obs}(1+z)^{\gamma-1}$, where the factor $(1+z)^{-1}$ accounts for cosmological time dilation and $(1+z)^\gamma$ accounts for the wavelength dependence of the dust lag. Following previous dust RM studies \citep{2023MNRAS.522.3439C,2024ApJ...968...59M,2025ApJ...993..203T,2025ApJ...994..188A}, we estimate the value of $\gamma$ for our sample using the relation $\gamma=\log(\tau_{\rm W2}/\tau_{\rm W1})/\log(\lambda_{\rm W2}/\lambda_{\rm W1})$. Based on the ICCF lag measurements of the core sample, we obtain a median lag ratio of $\tau_{\rm W2}/\tau_{\rm W1}=1.24^{+0.33}_{-0.21}$, corresponding to $\gamma\simeq0.71$. This value is broadly consistent with previous measurements, which typically yield $\tau_{\rm W2}/\tau_{\rm W1}\sim1.15-1.26$ \citep{2019ApJ...886...33L,2023MNRAS.522.3439C,2024ApJ...968...59M,2025ApJ...993..203T,2025ApJ...994..188A}. We therefore apply a redshift correction factor of $(1+z)^{-0.29}$ to the observed-frame lags, and use the corrected rest-frame lags in the subsequent analysis.

\subsection{Accretion Disk Contamination Correction}\label{subsec:Accretion Disk Contamination Correction}

The observed infrared flux in the W1 and W2 bands contains emission from both the dusty torus and the accretion disk. The ADC may lead to an underestimation of the intrinsic dust reverberation lag \citep{2024ApJ...968...59M,2025ApJ...993..203T}. Therefore, it is necessary to remove the ADC from the observed infrared light curves before deriving the intrinsic dust lags.

We estimate the accretion-disk contribution in the observed infrared light curves using the prescription
provided by \citet{2014ApJ...788..159K}, 
\begin{equation}
\label{eq:ADC_IR1} 
\begin{aligned} 
F_{\rm IR}^{\rm AD}(t)=F_{\rm opt}(t)\left(\frac{\nu_{\rm IR}}{\nu_{\rm opt}}\right)^{\alpha},
\end{aligned} 
\end{equation}
where $F_{\rm opt}(t)$ is the optical continuum flux, $\nu_{\rm opt}$ and $\nu_{\rm IR}$ are the effective frequencies of the optical and infrared bands, respectively, and $\alpha$ is the power-law index of the accretion-disk spectrum. Following previous dust RM studies \citep{2019ApJ...886...33L,2024ApJ...968...59M,2025ApJ...993..203T}, we adopt $\alpha=1/3$, as expected from the standard thin-disk model \citep{1973A&A....24..337S}. This choice is also supported by polarized near-IR spectra, which suggest that the intrinsic accretion-disk continuum can extend into the near-IR with a spectral slope close to $\alpha=1/3$ \citep{2008Natur.454..492K}. For each WISE epoch, we estimate $F_{\rm opt}(t)$ from the DRW reconstruction of the intercalibrated optical light curve, which provides the optical continuum flux at the W1 and W2 observation epochs. The dust-only infrared flux is then obtained by subtracting the estimated AD component from the observed infrared flux,
\begin{equation}
\label{eq:ADC_IR2} 
\begin{aligned} 
F_{\rm IR}^{\rm dust}(t)=F_{\rm IR}^{\rm obs}(t)-F_{\rm IR}^{\rm AD}(t).
\end{aligned} 
\end{equation}

After applying the ADC correction, we remeasured the W1 and W2 dust lags for the 65 sources in our core sample using the ICCF, JAVELIN, and PYROA methods. The same lag-searching windows and method-specific configurations were adopted to ensure a direct comparison with the non-ADC lag measurements. Figure~\ref{fig:Lag_comparison_ADC} compares the dust lags derived with and without ADC correction for the W1 and W2 bands and for each lag-measurement method. To quantify the effect of the correction, we fit the relation between the ADC-corrected and uncorrected dust radii ($R_{\rm ADC}$ and $R_{\rm non-ADC}$) using 

\begin{equation}
\label{eq:ADC_Compare} 
\begin{aligned} 
\log\frac{R_{\rm ADC}}{100~{\rm light\text{-}day}}=k  \log \frac{R_{\rm non-ADC}}{100~{\rm light\text{-}day}}+b.
\end{aligned} 
\end{equation}
For all three methods, the ADC-corrected and uncorrected lags show a tight correlation, with best-fit slopes and intercepts reported in the corresponding panels of Figure~\ref{fig:Lag_comparison_ADC}. The fitted slopes are close to unity in both the W1 and W2 bands, while the intercepts in the W1 band are systematically larger than those in the W2 band. This suggests that the ADC correction has a more pronounced effect in the W1 band, which is consistent with the expectation that the accretion-disk contribution becomes stronger at shorter infrared wavelengths. Finally, the W1 and W2 dust lag measurements for all sources in the core sample, derived using the ICCF, JAVELIN, and PYROA methods with and without ADC correction, are summarized in Table~\ref{tab:lag}.

\begin{figure*}[t]
    \centering
    \includegraphics[width=0.90\textwidth]{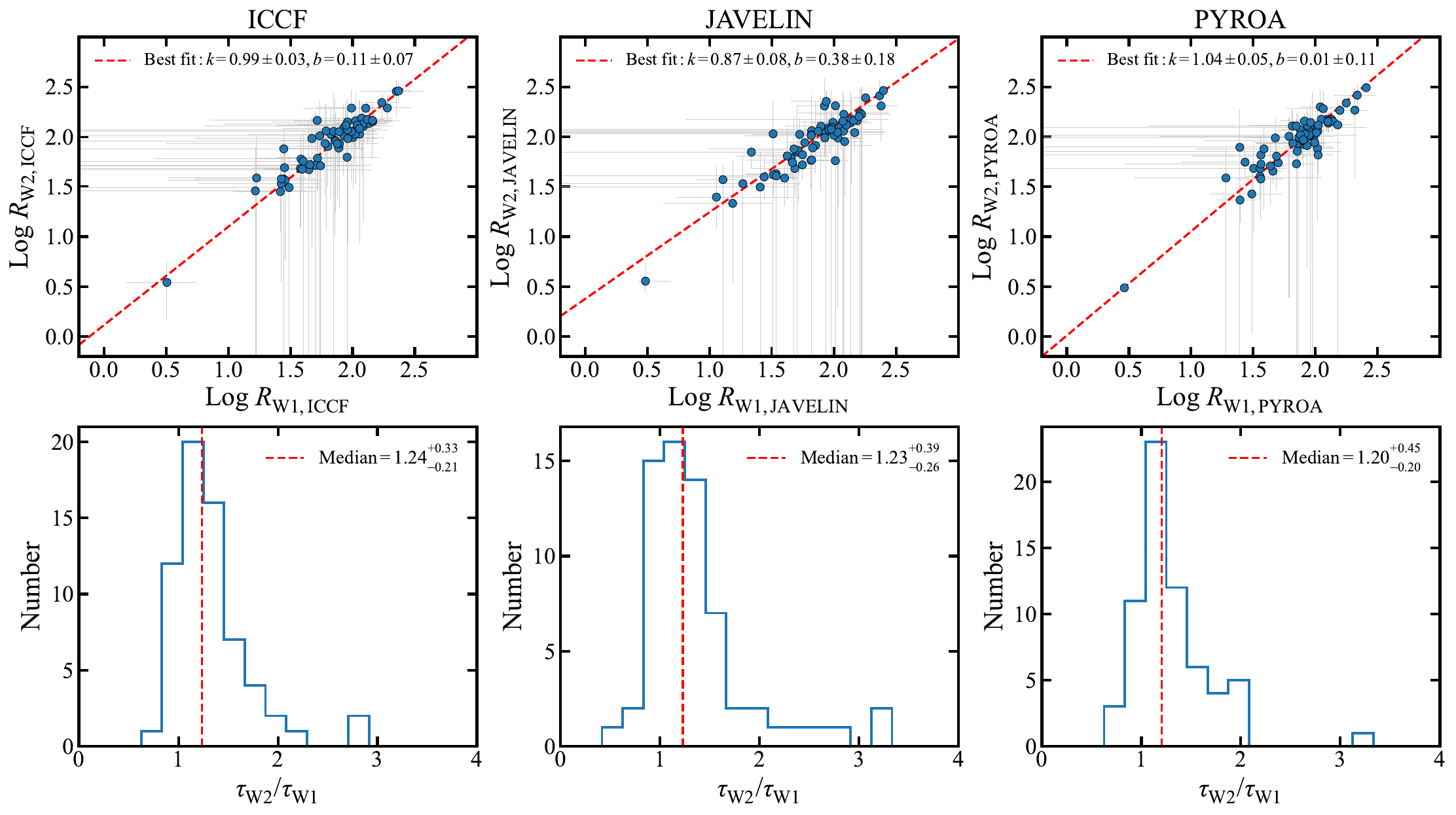} 
    \caption{Comparison of the ADC-corrected W1- and W2-band lags. The top row compares the W1- and W2-band lags derived from ICCF (left), JAVELIN (middle), and PYROA (right), with the red dashed line indicating the best-fit relation. The bottom row shows the distributions of the lag ratio ($\tau_{\rm ratio}=\tau_{\rm W2}/\tau_{\rm W1}$) for the three methods, where the red dashed line marks the median value.}
    \label{fig:Lag_comparison_W1W2}
\end{figure*}

\section{Results and Discussion} \label{sec:discussion}

\subsection{Comparison of Lags from Different Methods}\label{subsec:Comparison of Different Methods}

To assess the reliability of the lag measurements in our core sample, we compare the ADC-corrected observed-frame dust lags obtained from the ICCF, JAVELIN, and PYROA methods in Figure~\ref{fig:Lag_comparison_methods}. Overall, the lag measurements from the three methods show good agreement in both the W1 and W2 bands, with no significant systematic offset from the 1:1 relation. To quantify the consistency, we calculate the scatter and median offset in logarithmic lag space for each pair of methods. In the W1 band, the JAVELIN-to-ICCF comparison shows the best agreement (scatter = 0.13 dex, offset = 0.004 dex), while the PYROA-to-ICCF and PYROA-to-JAVELIN comparisons exhibit slightly larger scatters (0.14--0.17 dex) and offsets (\(<0.03\) dex). Similar consistency is found in the W2 band, with scatters of 0.11--0.14 dex and offsets below 0.02 dex among different method pairs. These small scatters and negligible offsets demonstrate that the lag measurements are largely insensitive to the choice of lag-measurement method.

\subsection{Comparison of Torus Sizes in W1 and W2 Bands}\label{subsec:Comparison of W1 and W2}
The W1 (3.4\,$\mu$m) and W2 (4.6\,$\mu$m) bands probe dust emission at different temperatures. In the stratified dusty torus model \citep{2008ApJ...685..160N,2012MNRAS.420..526M}, hotter dust traced by the W1 band resides closer to the central engine than cooler dust traced by the W2 band. Thus, comparing the torus sizes measured in these two bands directly tests this spatial and temperature stratification.

In the top row of Figure \ref{fig:Lag_comparison_W1W2}, we compare the ADC-corrected lags in the W2 band against those in the W1 band for the core sample. By performing a linear regression in logarithmic space, we obtain the best-fit relation for the primary ICCF measurements:
\begin{equation} \label{eq:W1_W2_relation} 
\log R_{\rm W2} = (0.99 \pm 0.03) \log R_{\rm W1} + (0.11 \pm 0.07). 
\end{equation}
Although the best-fit slopes and intercepts exhibit slight variations when adopting the JAVELIN and PYROA measurements, all three methods consistently demonstrate a robust positive correlation where the W2-band lags are systematically longer than the W1-band lags in our sample. 

To quantify the relative difference between the W1 and W2 emitting regions, we investigate the distribution of the lag ratio, defined as $\tau_{\rm ratio} = \tau_{\rm W2}/\tau_{\rm W1}$. As illustrated in the bottom row of Figure \ref{fig:Lag_comparison_W1W2}, the lag-ratio distributions for all three methods are centered above unity. The median lag ratios are $1.24^{+0.33}_{-0.21}$ (ICCF), $1.23^{+0.37}_{-0.26}$ (JAVELIN), and $1.20^{+0.45}_{-0.20}$ (PYROA), where the uncertainties correspond to the 16th and 84th percentiles of the distributions. We also examined the lag ratio distribution using the lag measurements without ADC correction. In this case, the median lag ratios are slightly higher, yielding $1.26^{+0.42}_{-0.19}$, $1.31^{+0.69}_{-0.25}$, and $1.32^{+0.49}_{-0.28}$ for the three respective methods.

Although these lag ratios are broadly consistent with previous dust RM studies within the uncertainties, they appear mildly larger than those obtained from higher-luminosity AGN samples (e.g., $\sim 1.15$ in \citealt{2019ApJ...886...33L}; $\sim 1.19$ in \citealt{2024ApJ...968...59M}; $\sim 1.21$ in \citealt{2025ApJ...993..203T}). To further explore whether this difference is related to AGN properties, we combine our measurements with those of \citet{2024ApJ...968...59M} and examine the dependence of the lag ratio on bolometric luminosity and Eddington ratio (see Figure~\ref{fig:R-Lag_ratio_dependence}). We find a weak anti-correlation between the lag ratio and bolometric luminosity ($\rho=-0.24$, $p=3.1\times10^{-7}$), while the anti-correlation with Eddington ratio is weaker ($\rho=-0.18$, $p=1.1\times10^{-3}$). 

\begin{figure*}[t]
    \centering
    \includegraphics[width=0.90\textwidth]{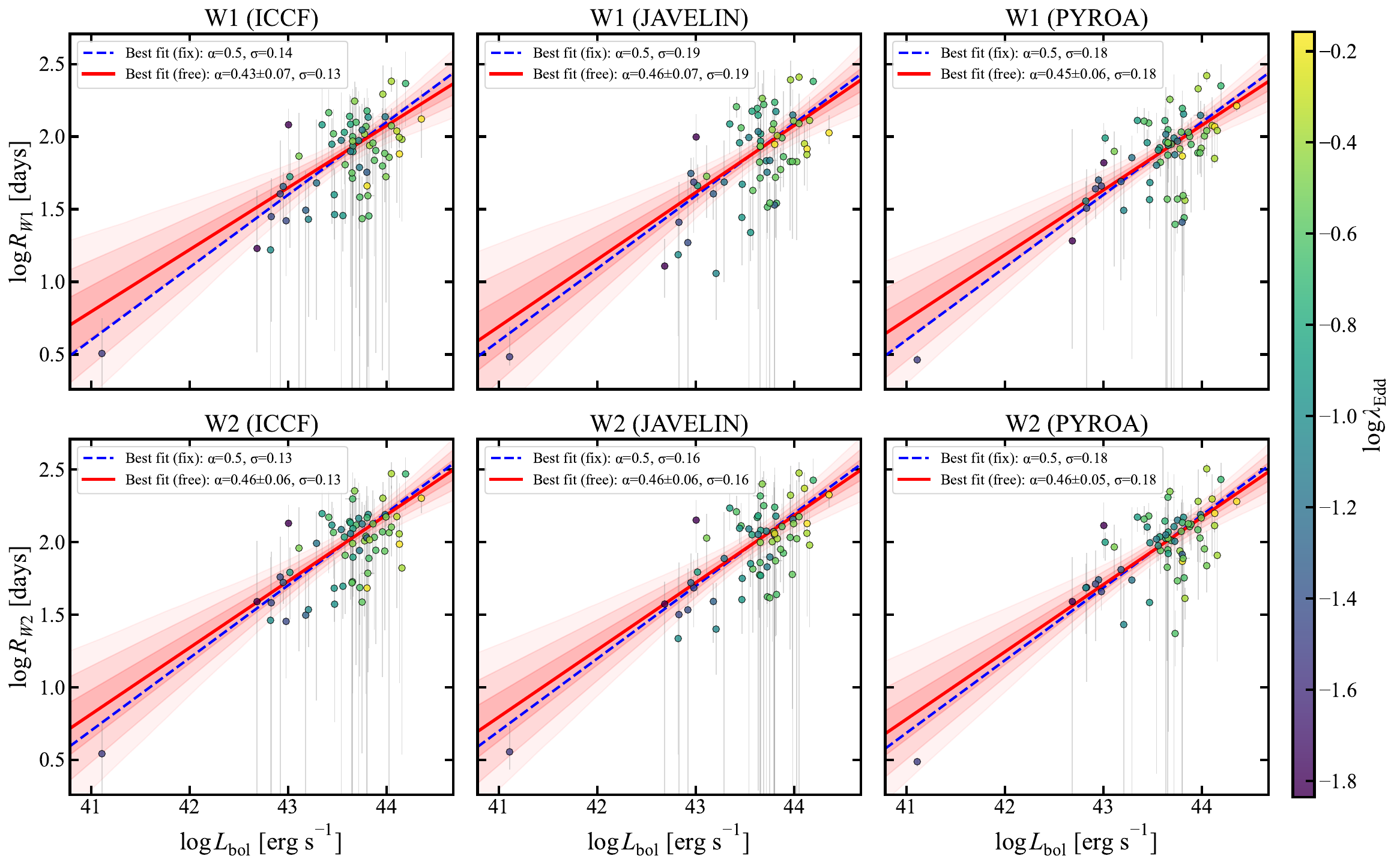} 
    \caption{The $R_{\rm dust}-L_{\rm AGN}$ relation for low-mass AGNs. The top and bottom rows show the results for the W1 and W2 bands, respectively. The left, middle, and right columns present the measurements derived from ICCF, JAVELIN, and PYROA, respectively. The data points are color-coded by the Eddington ratio. The blue dashed lines represent the best-fit regression with a fixed slope of 0.5, while the red solid lines show the best-fit regression with a free slope. The red shaded regions indicate the 1$\sigma$, 2$\sigma$, and 3$\sigma$ confidence intervals around the best-fit regression line.}
    \label{fig:R_L_relation_logL_bol_ADC}
\end{figure*}

This trend is consistent with the luminosity-dependent dust structure scenario suggested by \citet{2024ApJ...968...59M}. As illustrated in Figure 13 of \citet{2024ApJ...968...59M}, high-luminosity AGNs exhibit a compact dust emission region where the W1 and W2 emitting regions become closer. Conversely, as luminosity decreases, the spatial separation between the hotter (W1) and cooler (W2) dust emission regions becomes more pronounced, naturally leading to larger lag ratios. With high-cadence Spitzer monitoring, \citet{2015ApJ...801..127V} reported dust lags of $\tau_{3.6\,\mu\text{m}} = 37.2_{-2.2}^{+2.4}$ days and $\tau_{4.5\,\mu\text{m}} = 47.1_{-3.1}^{+3.1}$ days for NGC 6418 ($L_{\rm bol} \sim 10^{43.3} \text{erg s}^{-1}$), yielding a lag ratio of $\sim1.266$. The lag ratio of this low-luminosity AGN is higher than the typical lag ratios reported for the high-luminosity AGN sample in \citet{2024ApJ...968...59M}, and is comparable to the typical values measured for our low-luminosity AGN sample.

In summary, these findings suggest that the temperature stratification of the dusty torus appears to be a common feature across a wide range of black hole masses and luminosities. However, compared to the more compact emission regions found in high-luminosity AGNs, the sources in our sample may host relatively more extended and distinctly stratified structures, indicating a luminosity-dependent modulation of the torus geometry.

\subsection{Dust Radius–Luminosity Relation for Low-Mass AGNs}\label{subsec:RL Relation}
We investigate the $R_{\rm dust}-L_{\rm AGN}$ relation using our core sample of 65 low-mass AGNs. To parameterize the $R_{\rm dust}-L_{\rm AGN}$ relation, we perform a linear regression in the logarithmic space using the following form:
\begin{equation} \label{eq:RL_relation}
\log \left( \frac{R_{\rm dust}}{1\,\text{light-day}} \right) = \beta + \alpha \log \left( \frac{L_{\rm AGN}}{10^{43}\,\text{erg s}^{-1}} \right). 
\end{equation}
We perform the linear regression using the \texttt{emcee} package\footnote{\url{https://emcee.readthedocs.io/en/stable/tutorials/line/}}
\citep{2013PASP..125..306F}, with a likelihood function that accounts for
the measurement uncertainties in both variables and includes an intrinsic
scatter term for the correlation. We consider two fitting scenarios: one with $\alpha$ as a free parameter and another with $\alpha=0.5$ fixed, as expected from dust radiative equilibrium \citep{1987ApJ...320..537B}.

We adopt the rest-frame, ADC-corrected lag measurements from ICCF, JAVELIN, and PYROA to estimate the torus size ($R_{\rm dust}=c\tau_{\rm dust}$). We first use the bolometric luminosity ($L_{\rm bol}$) as the proxy for $L_{\rm AGN}$. The corresponding best-fit relations for the W1 and W2 bands are illustrated in Figure \ref{fig:R_L_relation_logL_bol_ADC}. For the ADC-corrected ICCF lag measurements, the best-fit free-slope relations are:
\begin{align}
\log R_{\mathrm{W1,ICCF}}
&= (1.65 \pm 0.06) \notag \\
& + (0.43 \pm 0.07)
\log \left( \frac{L_{\mathrm{bol}}}{10^{43}\,\mathrm{erg\,s^{-1}}} \right),
\label{eq:W1_bol_fit}
\\[0.6em]
\log R_{\mathrm{W2,ICCF}}
&= (1.73 \pm 0.05) \notag \\
& + (0.46 \pm 0.06)
\log \left( \frac{L_{\mathrm{bol}}}{10^{43}\,\mathrm{erg\,s^{-1}}} \right).
\label{eq:W2_bol_fit}
\end{align}
These results demonstrate that our low-mass AGN sample follows tight $R_{\rm dust}-L_{\rm AGN}$ relations with small intrinsic scatters of $\sigma_{\mathrm{free}} = 0.13 \pm 0.03$ dex for both the W1 and W2 bands. The observed slopes ($\alpha = 0.43 \pm 0.07$ for W1 and $0.46 \pm 0.06$ for W2) are slightly shallower than the theoretical expectation of 0.5. Additionally, the alternative lag measurements derived from JAVELIN and PYROA yield consistent $R_{\rm dust}-L_{\rm AGN}$ relations, with best-fit free slopes remaining below 0.5 for both the W1 and W2 bands. However, these $R_{\rm dust}-L_{\rm AGN}$ relations based on the JAVELIN and PYROA methods exhibit slightly larger intrinsic scatters ($\sigma_{\mathrm{free}} \sim 0.16 - 0.19\,\mathrm{dex}$) compared to those anchored by the ICCF method. As a robust check of our fitting method, we also implemented the \texttt{LINMIX} \citep{2007ApJ...665.1489K} and Bivariate Correlated Errors and intrinsic Scatter (\texttt{BCES}; \citealt{1996ApJ...470..706A}) linear regression codes. Both estimators deliver consistent fitting results, supporting the robustness of the derived slopes, which are mildly shallower than 0.5.

To further examine the dependence of the $R_{\rm dust}-L_{\rm AGN}$ relations on the choice of AGN luminosity proxy, we substitute $L_{\mathrm{bol}}$ with alternative luminosity indicators. The catalog of \citet{2026HXG} also provides the total monochromatic luminosity at 5100\,\AA\ ($L_{5100}$) and the host-galaxy-subtracted AGN luminosity at 5100\,\AA\ ($L_{5100,\mathrm{AGN}}$). In addition, we explore the $R_{\rm dust}-L_{\rm AGN}$ relations using the simultaneous infrared continuum luminosities ($L_{\rm W1}$ and $L_{\rm W2}$), derived from the average fluxes of the ADC-corrected WISE light curves. The $R_{\rm dust}-L_{\rm AGN}$ relations and linear regression plots for these alternative luminosity proxies are displayed in Figure \ref{fig:R-L_relations}. Furthermore, we repeat the analysis without applying the ADC correction to evaluate its impact on the inferred $R_{\rm dust}-L_{\rm AGN}$ relations. The complete grid of the best-fit regression parameters across all combinations of lag methods and luminosity proxies is comprehensively summarized in Table \ref{tab:R-L Relations}.

Based on the regression analysis detailed in Table \ref{tab:R-L Relations}, several key characteristics governing the $R_{\rm dust}-L_{\rm AGN}$ scaling relations for our low-mass AGN sample can be summarized as follows: (1) The best-fit free slopes ($\alpha_{\mathrm{free}}$) of the $R_{\rm dust}-L_{\rm AGN}$ relations generally range from $\sim 0.30$ to $0.50$, depending on the specific choices of the lag-measurement method and AGN luminosity proxy. All best-fit slopes are below the theoretical expectation of 0.5, although some are consistent with this value within uncertainties. (2) The $R_{\rm dust}-L_{\rm AGN}$ relations constructed from the three different lag-measurement methods (ICCF, JAVELIN, and PYROA) exhibit generally consistent regression parameters. Among these, the relations anchored by the ICCF method consistently show the smallest intrinsic scatter.
(3) Applying the ADC correction does not significantly change the best-fit slopes of the $R_{\rm dust}-L_{\rm AGN}$ relations, but generally increases the fitted intercepts, with a stronger effect in the W1 band than in the W2 band. In addition, the $R_{\rm dust}-L_{\rm AGN}$ relations with ADC correction generally show smaller intrinsic scatters.
(4) Compared with the bolometric luminosity and optical luminosity proxies (i.e., $L_{\rm bol}$, $L_{5100}$, and $L_{5100,\mathrm{AGN}}$), employing the contemporaneous infrared luminosities ($L_{\rm W1}$ and $L_{\rm W2}$) as proxies for $L_{\rm AGN}$ yields tighter $R_{\rm dust}-L_{\rm AGN}$ relations, bringing the intrinsic scatter down to $\sim 0.1\,\mathrm{dex}$ for both the W1 and W2 bands.

\begin{figure*}[t]
    \centering
    \includegraphics[width=0.90\textwidth]{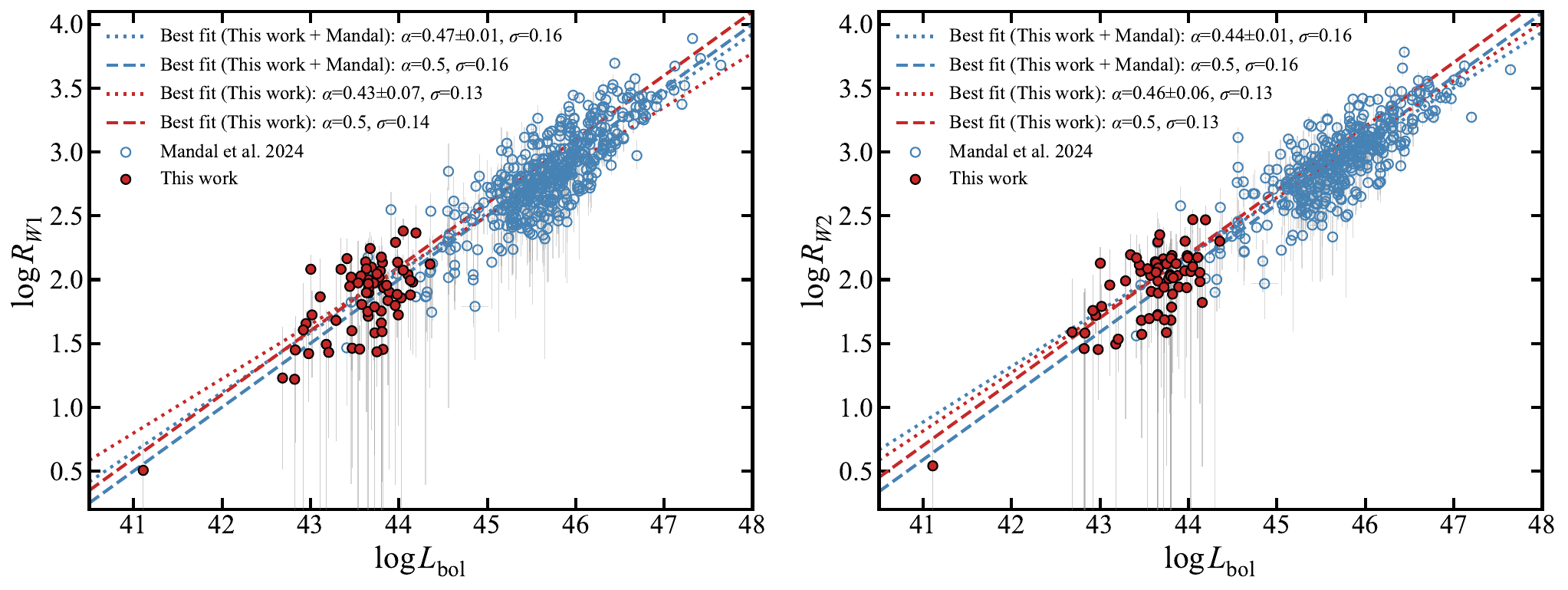} 
    \caption{The $R_{\rm dust}-L_{\rm AGN}$ relation for the combined sample of this work and \citet{2024ApJ...968...59M}. The left and right panels show the W1- and W2-band relations based on ICCF lags, respectively. The red filled circles represent the sample in this work, while the blue open circles denote the sample from \citet{2024ApJ...968...59M}. The dashed and dotted lines show the best-fit relations with fixed and free slopes, respectively. The red and blue lines represent the fits for the sample in this work and the combined sample of this work and \citet{2024ApJ...968...59M}, respectively.}
    \label{fig:R-L_relation_Merged_ICCF_ADC}
\end{figure*}

Overall, the $R_{\rm dust}-L_{\rm AGN}$ relations for our low-mass AGN sample are broadly consistent with previous WISE-based dust RM studies, which generally found best-fit slopes shallower than the theoretical expectation of 0.5. Earlier work on luminous PG quasars reported slopes of $\sim0.45-0.47$ \citep{2019ApJ...886...33L}, while several more recent WISE-based studies found slopes of $\sim0.33-0.41$ \citep{2023MNRAS.522.3439C,2024ApJ...968...59M,2025ApJ...993..203T,2025ApJ...994..188A}. 
The shallow slopes may be partly related to the choice of AGN luminosity proxy. As shown in Table~\ref{tab:R-L Relations}, the best-fit slopes of the $R_{\rm dust}-L_{\rm AGN}$ relations vary systematically with the adopted luminosity indicators, ranging from $\sim0.3$ to $0.5$. This dependence suggests that different luminosity proxies may differ in their ability to represent the AGN emission that governs dust heating. In particular, monochromatic or single-band luminosities may not fully capture the intrinsic luminosity responsible for dust heating, as they can be affected by luminosity-dependent variations in the AGN SEDs. Similar effects have been reported by \citet{2024A&A...690A..76G}, who compared the $R_{\rm dust}-L_{\rm AGN}$ relations using different luminosity indicators, including X-ray, optical, mid-infrared monochromatic luminosities, as well as bolometric luminosities derived from these indicators. They found that the best-fit slopes depend on the adopted luminosity proxy, ranging from $0.37$ to $0.50$. In addition to luminosity-proxy effects, the accretion state may also impact the observed slope of the $R_{\rm dust}-L_{\rm AGN}$ relation. In high-Eddington-ratio AGNs, the self-shadowing effect of slim disks can reduce the effective illumination of the dusty torus, leading to shorter dust lags and a flatter observed $R_{\rm dust}-L_{\rm AGN}$ relation \citep{2011ApJ...737..105K,2014ApJ...797...65W,2018ApJ...856....6D}. Compared with some previous dust RM studies of luminous AGN samples, our low-mass AGN sample contains a lower fraction of high-Eddington-ratio sources, as shown in the right panel of Figure~\ref{fig:distribution}. This suggests that the impact of the self-shadowing effect may be weaker in our sample, which could partly explain why the slopes of the $R_{\rm dust}-L_{\rm bol}$ relations are closer to the theoretical expectation of 0.5. A more detailed discussion of the Eddington-ratio dependence is presented in Section~\ref{subsec:Deviation}.

\begin{figure*}[t]
    \centering
    \includegraphics[width=0.90\textwidth]{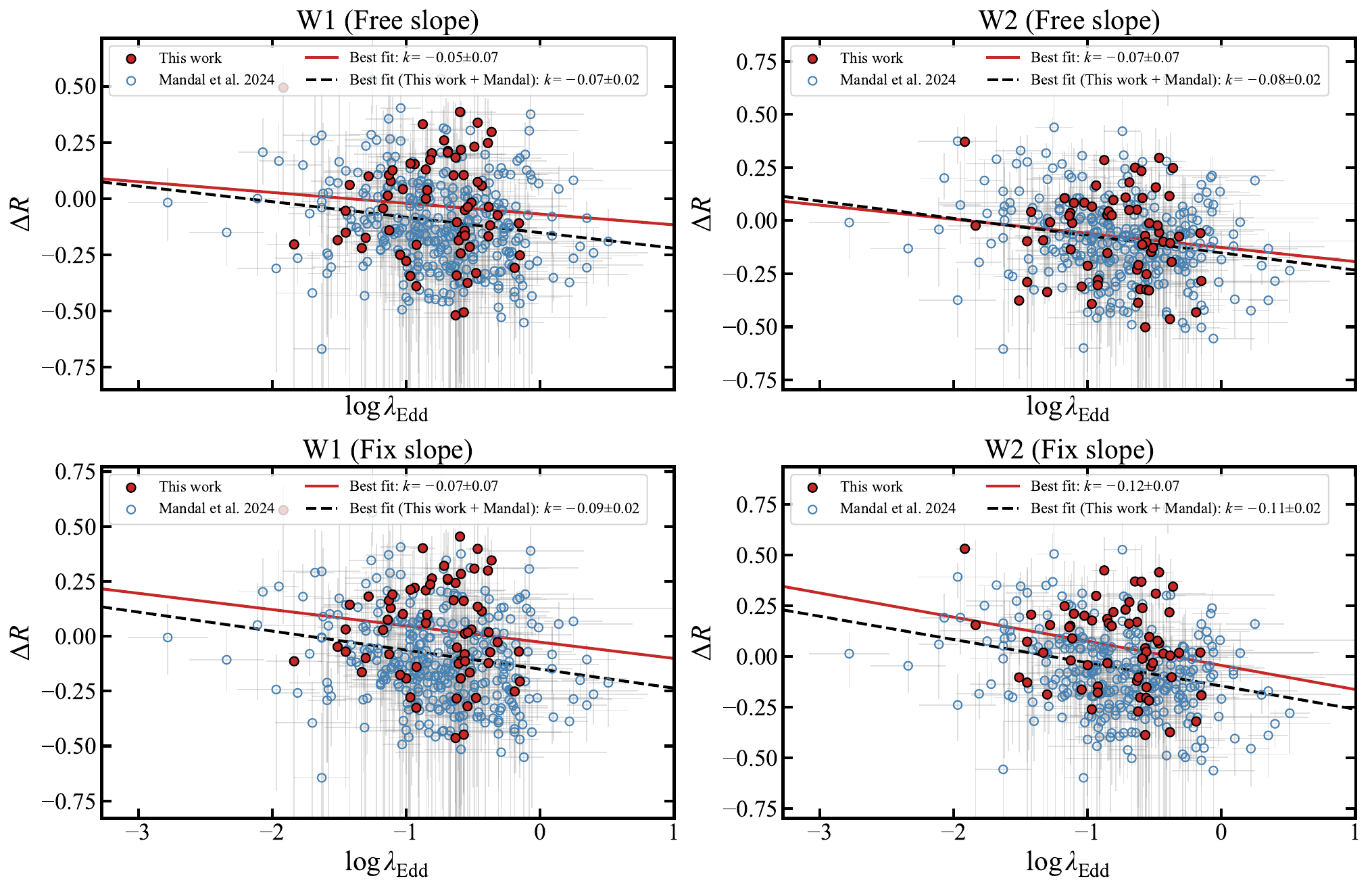} 
    \caption{Correlations of the lag deviations and the Eddington ratio. The top-left and top-right panels show the correlations of $\Delta R_{\mathrm{W1,free}}$ and $\Delta R_{\mathrm{W2,free}}$ with $\log \lambda_{\mathrm{Edd}}$, respectively. The bottom-left and bottom-right panels show the correlations of $\Delta R_{\mathrm{W1,fix}}$ and $\Delta R_{\mathrm{W2,fix}}$ with $\log \lambda_{\mathrm{Edd}}$, respectively. The red filled circles represent data from this work, and the blue open circles denote data from \citet{2024ApJ...968...59M}. The red solid line shows the best-fit relation for the sample in this work, while the black dashed line shows the best-fit relation for the combined sample of this work and \citet{2024ApJ...968...59M}.}
    \label{fig:Deviation_Edd}
\end{figure*}

\subsection{Joint Analysis with Previous Dust RM Samples}\label{subsec:Joint Analysis}
\subsubsection{A Unified Dust Radius--Luminosity Relation across Luminosity Regimes}\label{subsubsect:UnifiedRL}
To better constrain the $R_{\rm dust}-L_{\rm AGN}$ relation over a broader luminosity range, we perform a joint analysis by combining our low-luminosity sample with the higher-luminosity AGN sample of \citet{2024ApJ...968...59M}. Their systematic analysis provides W1- and W2-band lag measurements both before and after ADC correction, enabling a direct comparison with our results.

We perform a joint linear regression on the combined sample using the primary ICCF lag measurements and bolometric luminosities. The corresponding joint $R_{\rm dust}-L_{\rm bol}$ relations are shown in Figure~\ref{fig:R-L_relation_Merged_ICCF_ADC}. Our low-mass AGN sample extends dust RM measurements toward the low-luminosity regime and provides important constraints at the low-luminosity end of the relations. After applying the ADC correction, the best-fit $R_{\rm dust}-L_{\rm bol}$ relations for the combined sample are
\begin{align}
\log  R_{\mathrm{W1,ICCF}}
&= (2.05 \pm 0.02) \notag \\
& + (0.47 \pm 0.01)
\log \left( \frac{L_{\mathrm{bol}}}{10^{44}~{\rm erg~s^{-1}}} \right),
\label{eq:W1_merged_fit}
\\[0.6em]
\log R_{\mathrm{W2,ICCF}}
&= (2.19 \pm 0.02) \notag \\
& + (0.44 \pm 0.01)
\log \left( \frac{L_{\mathrm{bol}}}{10^{44}~{\rm erg~s^{-1}}} \right).
\label{eq:W2_merged_fit}
\end{align}
We also perform fixed-slope fits ($\alpha=0.5$) and repeat the joint regression using the lag measurements without ADC correction. The full set of best-fit parameters for these joint analyses is summarized in Table~\ref{tab:R-L Relations}.

The combined $R_{\rm dust}-L_{\rm bol}$ relations yield slopes of 0.47 $\pm$ 0.01 for W1 and 0.44 $\pm$ 0.01 for W2, with intrinsic scatters of 0.16 $\pm$ 0.01 dex in both bands. These slopes are closer to the theoretical expectation of 0.5 than those obtained by \citet{2024ApJ...968...59M}, which are 0.39 for W1 and 0.33 for W2. This suggests that extending dust RM measurements to low-mass AGNs provides important leverage for anchoring the low-luminosity end of the $R_{\rm dust}-L_{\rm AGN}$ relation and constraining a scaling relation more consistent with the dust radiative equilibrium prediction.

\begin{table*}[t]
    \begin{center}
    \caption{Correlations between the Lag Deviations and Eddington Ratio.}
    \vspace{-1em}
    \label{tab:eddington_relation}
    \setlength\tabcolsep{1.2mm}{
    \begin{tabular}{ccccccccc}
        \toprule
        \toprule
        \multirow{2}{*}{Case} & \multicolumn{4}{c}{This work} & \multicolumn{4}{c}{This work + \citet{2024ApJ...968...59M}} \\
        & $k$ & $b$ & $r$ & $p$ & $k$ & $b$ & $r$ & $p$ \\
        (1)&(2)&(3)&(4)&(5)&(6)&(7)&(8)&(9) \\
        \midrule
        $\Delta R_{\mathrm{W1, free}}$ vs.\ log $\lambda_{\mathrm{Edd}}$
        & $-0.05 \pm 0.07$ & $-0.07 \pm 0.06$ & $-0.04$ & $7.81\times10^{-1}$
        & $-0.07 \pm 0.02$ & $-0.15 \pm 0.02$ & $-0.16$ & $1.51\times10^{-3}$ \\

        $\Delta R_{\mathrm{W2, free}}$ vs.\ log $\lambda_{\mathrm{Edd}}$
        & $-0.07 \pm 0.07$ & $-0.12 \pm 0.06$ & $-0.11$ & $4.02\times10^{-1}$
        & $-0.08 \pm 0.02$ & $-0.15 \pm 0.02$ & $-0.19$ & $2.58\times10^{-4}$ \\

        $\Delta R_{\mathrm{W1, fix}}$ vs.\ log $\lambda_{\mathrm{Edd}}$
        & $-0.07 \pm 0.07$ & $-0.02 \pm 0.06$ & $-0.09$ & $4.90\times10^{-1}$
        & $-0.09 \pm 0.02$ & $-0.15 \pm 0.02$ & $-0.19$ & $1.48\times10^{-4}$ \\

        $\Delta R_{\mathrm{W2, fix}}$ vs.\ log $\lambda_{\mathrm{Edd}}$
        & $-0.12 \pm 0.07$ & $-0.04 \pm 0.06$ & $-0.18$ & $1.61\times10^{-1}$
        & $-0.11 \pm 0.02$ & $-0.15 \pm 0.02$ & $-0.26$ & $1.29\times10^{-6}$ \\
        \bottomrule
    \end{tabular}}
    \end{center}
    \textbf{Note.} Column (1): Different cases used in the fitting. 
Columns (2)–(5): Fitting results for the sample in this work, including the slope ($k$), intercept ($b$), Spearman rank correlation coefficient ($r$) and probability value of no correlation ($p$). 
Columns (6)–(9): Fitting results for the combined sample of this work and \citet{2024ApJ...968...59M}, including the slope ($k$), intercept ($b$), Spearman rank correlation coefficient ($r$) and probability value of no correlation ($p$). 
\end{table*}

\subsubsection{Deviation from the Dust Radius–Luminosity Relation}\label{subsec:Deviation}

We further explore whether the deviation from the $R_{\rm dust}-L_{\rm AGN}$ relation is related to the accretion state of the AGN. For each source, we define the lag deviation as the logarithmic offset between the measured dust radius and the radius predicted by the best-fit $R_{\rm dust}-L_{\rm bol}$ relation, $\Delta R = \log R_{\rm obs} - \log R_{\rm pred}$. Here, $R_{\rm obs}$ is the dust radius measured from the ADC-corrected ICCF lag, and $R_{\rm pred}$ is the predicted radius from the best-fit $R_{\rm dust}-L_{\rm bol}$ relation. We calculate the deviations separately for the W1 and W2 bands. In addition, we consider two definitions of $R_{\rm pred}$: one based on the free-slope $R_{\rm dust}-L_{\rm bol}$ relation and the other based on the relation with the slope fixed to the theoretical value of 0.5. These correspond to $\Delta R_{\rm W1,free}$, $\Delta R_{\rm W2,free}$, $\Delta R_{\rm W1,fix}$, and $\Delta R_{\rm W2,fix}$.

We perform linear regression analysis to characterize the relationship between the lag deviations and $\log \lambda_{\rm Edd}$ for both our low-mass AGN sample alone and the combined sample of this work and \citet{2024ApJ...968...59M}. For each case, we calculate the best-fit slope ($k$), intercept ($b$), the Spearman rank correlation coefficient ($r$), and the corresponding null-hypothesis probability ($p$). The fitting results are summarized in Table~\ref{tab:eddington_relation}, and the correlations are shown in Figure~\ref{fig:Deviation_Edd}. For our low-mass AGN sample, the lag deviations across all cases exhibit weak negative correlations with $\log \lambda_{\rm Edd}$, yielding slopes and correlation coefficients below zero ($k < 0$ and $r < 0$). However, these trends are statistically insignificant, with all corresponding $p$-values exceeding the significance threshold of 0.1. In contrast, the combined sample exhibits stronger and statistically significant negative correlations, with all corresponding $p$-values below 0.002.

Similar negative correlations between the lag deviations and Eddington ratio have been reported in recent dust RM studies \citep{2023MNRAS.522.3439C,2024ApJ...968...59M,2025ApJ...993..203T,2025ApJ...994..188A}. This trend is physically consistent with the self-shadowing effect of slim disks. In high-Eddington-ratio AGNs, the inner accretion disk can become geometrically thick and partially shield the dusty torus from the central dust-heating radiation \citep{2011ApJ...737..105K,2014ApJ...797...65W,2018ApJ...856....6D}, leading to shorter dust lags at fixed luminosity and thus contributing to the observed flattening of the $R_{\rm dust}-L_{\rm AGN}$ relation.

A more detailed investigation was presented by \citet{2025ApJ...993..203T}, who analyzed an enlarged dust RM sample by combining their sample with that of \citet{2024ApJ...968...59M}. They found a negative correlation between the lag deviations and Eddington ratio in the combined sample. After dividing the sources into different redshift and luminosity bins, they further showed that this negative correlation persists across redshift bins but becomes much weaker or disappears within luminosity bins. This behavior is consistent with our results: the correlation is weak and statistically insignificant when only our low-mass AGN sample is considered, while a weak but statistically significant negative correlation emerges after combining our sample with that of \citet{2024ApJ...968...59M}. This difference may also be partly related to the Eddington-ratio distributions of the two samples. Although both samples are dominated by relatively low-Eddington-ratio sources, the sample of \citet{2024ApJ...968...59M} includes a higher fraction of high-Eddington-ratio AGNs, for which the self-shadowing effect of slim disks may be more pronounced.

Overall, although the Eddington-ratio dependence of the $R_{\rm dust}-L_{\rm AGN}$ relation becomes more pronounced when a broader AGN luminosity range is considered, it remains weak within our low-mass AGN sample. Nevertheless, this weak dependence may still partly contribute to the mild deviation of the observed $R_{\rm dust}-L_{\rm AGN}$ relation from the theoretical $R \propto L^{0.5}$ scaling expected from dust radiative equilibrium.

\begin{figure*}[t]
    \centering
    \includegraphics[width=0.49\textwidth]{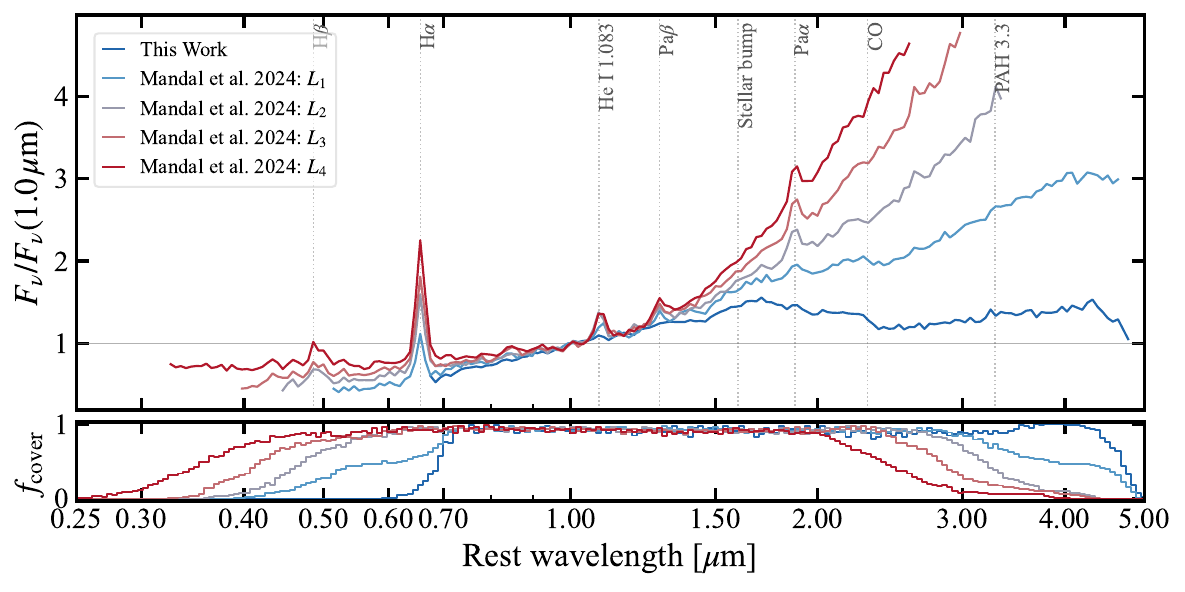}
    \hspace{-0.30cm}
    \includegraphics[width=0.49\textwidth]{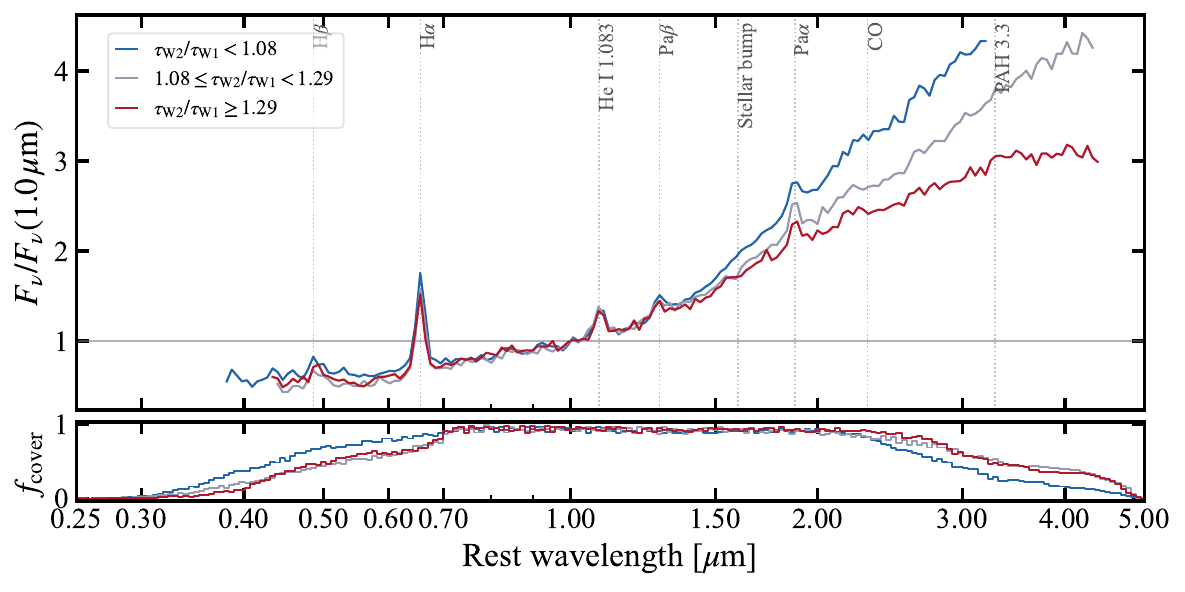}
    \vspace{-0.15cm}
    \caption{Optical-to-near-infrared composite spectra of AGNs constructed from SPHEREx spectrophotometric data. The left and right panels show the composite spectra constructed in different bolometric luminosity bins and W2-to-W1 lag-ratio bins, respectively. The corresponding bottom panels indicate the wavelength-dependent spectral coverage fraction.}
    \label{fig:composite_spectra}
\end{figure*}
\subsubsection{Optical-to-Near-Infrared Spectral Properties}\label{subsubsec:spec}
To further investigate the optical-to-near-infrared spectral properties of our low-mass AGN sample and compare them with the higher-luminosity AGN sample of \citet{2024ApJ...968...59M}, we utilize the SPHEREx spectrophotometric survey data, which provide all-sky near-infrared spectra over 0.75--5.0 $\mu$m with a spectral resolving power of $R\sim35$--$130$ \citep{2026ApJ...999..139B,2026ApJS..284...10H}. These data provide a unique opportunity to study the optical-to-near-infrared spectral variations of AGNs across different luminosity regimes and explore their possible connection with the dust temperature stratification inferred from RM measurements.

The SPHEREx spectrophotometric measurements are obtained using the \texttt{spxquery} package\footnote{\url{https://github.com/WenkeRen/spxquery}}. We adopt a quality-control procedure similar to that described by \citet{2026arXiv260902167R}, excluding measurements with non-finite values, unreliable quality flags, and significant outliers. We correct the measurements for Galactic extinction using the reddening map of \citet{1998ApJ...500..525S}, recalibrated by \citet{2011ApJ...737..103S}, and a standard Galactic extinction law with $R_V=3.1$. After the extinction correction, the spectra are shifted to the rest frame using the spectroscopic redshifts and interpolated onto a common logarithmic wavelength grid covering 0.25--5.0 $\mu$m with a logarithmic sampling interval of $\Delta\ln\lambda=0.015$. To minimize the effect of intrinsic luminosity differences, individual spectra are normalized using an iterative procedure based on spectral overlap. Following \citet{2026arXiv260710282K}, we first sort the targets by redshift and use an initial pair of overlapping spectra to establish a baseline composite. Each subsequent spectrum is normalized against the existing composite over the overlapping wavelength range, and the composite is updated as new spectra are added. The final composite spectrum is constructed by taking the median flux of the normalized spectra in each wavelength bin. The composite spectrum is then normalized to unity at 1 $\mu$m. The wavelength-dependent spectral coverage fraction ($f_{\rm cover}$), defined as the fraction of sources contributing to each wavelength bin, is recorded to assess the sampling completeness of the composite spectra. To avoid displaying poorly sampled spectral regions, we mask the portions of the composite spectra where $f_{\rm cover} < 0.3$.

We first investigate the luminosity dependence of the optical-to-near-infrared SEDs by constructing composite spectra in different bolometric luminosity bins. To ensure comparable statistical significance among different subsamples, the higher-luminosity AGN sample of \citet{2024ApJ...968...59M} is divided into four approximately equal-sized $L_{\rm bol}$ bins ($L_1-L_4$), with increasing luminosity from $L_1$ to $L_4$, while our low-mass AGN sample is treated as a separate low-luminosity group. As shown in the left panel of Figure~\ref{fig:composite_spectra}, the normalized optical-to-near-infrared composite spectra exhibit systematic variations with bolometric luminosity. The higher-luminosity composites show stronger optical-to-near-infrared continua, with a more pronounced near-infrared excess at $\sim1.5-3~\mu{\rm m}$. In contrast, our low-mass AGN sample exhibits a weaker near-infrared continuum excess, together with possible signatures of a stellar bump around 1.6~$\mu$m and CO absorption around 2.3~$\mu$m. These spectral differences may reflect luminosity-dependent variations in the relative contribution of hot-dust emission. However, the observed differences may also be affected by host-galaxy contamination, particularly for our low-mass AGN sample. Such luminosity-dependent SED variations may partly contribute to the differences in the $R_{\rm dust}-L_{\rm AGN}$ relations derived using different luminosity proxies, since monochromatic luminosities may not fully trace the intrinsic dust-heating luminosity.

We further investigate the connection between the optical-to-near-infrared SEDs and the dust temperature stratification inferred from the W2-to-W1 lag ratio. We combine our low-mass AGN sample with the higher-luminosity AGN sample of \citet{2024ApJ...968...59M} and divide the combined sample into three approximately equal-sized bins according to $\tau_{\rm W2}/\tau_{\rm W1}$, with boundaries at 1.08 and 1.29. Given the relatively large uncertainties in the individual lag ratios, we adopt this relatively coarse binning to explore possible trends in the composite SEDs. As shown in the right panel of Figure~\ref{fig:composite_spectra}, the composite spectra show no clear systematic differences below $\sim1~\mu$m, with similar optical continua and prominent H$\alpha$ and H$\beta$ emission features among the three bins. In contrast, systematic differences appear in the near-infrared continuum at $\sim1.5$--$3~\mu$m. The near-infrared excess relative to the normalized 1 $\mu$m flux becomes weaker with increasing lag ratio. Radiative-transfer models of clumpy AGN tori show that more centrally concentrated dust distributions produce relatively stronger near-infrared emission \citep{2010A&A...523A..27H}. This provides a possible qualitative interpretation of the observed trend, suggesting that sources with smaller W2-to-W1 lag ratios may have more centrally concentrated dust distributions and exhibit stronger near-infrared excess emission. However, the observed spectral differences may also be affected by luminosity-dependent SED variations and host-galaxy contributions. 

\subsection{Dust RM in a TDE Candidate}\label{subsec:TDE}
In our sample, we find J104330.05+512128.9 associated with the nuclear transient AT2019dqv, which was discovered by the ZTF in 2019 \citep{2019TNSTR.615....1N}. The transient has been interpreted as a TDE-like nuclear accretion flare based on its optical outburst, characterized by a rise timescale of $\sim67$ days and a much slower decay timescale of $\sim475$ days, accompanied by a strong infrared excess consistent with dust reprocessing \citep{2024MNRAS.529.2559V}. It was also classified as a dust-echo-like infrared flare by \citet{2025A&A...695A.228N}, owing to its significant infrared brightening and subsequent decline. The source has a black hole mass of $\log(M_{\rm BH}/M_\odot)=6.04$ and a bolometric luminosity of $\log (L_{\rm bol}/\mathrm{erg\,s}^{-1})=43.67$. Given its low black hole mass and the occurrence of a TDE-like nuclear transient, this source provides an interesting opportunity to examine dust RM following a transient nuclear outburst.

The source exhibits unusually long dust reverberation lags in both the W1 and W2 bands. As shown in Figure~\ref{fig:result_25}, the three independent methods yield generally consistent measurements, with $\tau_{\rm W1}=172.0^{+61.7}_{-49.4}$~days and $\tau_{\rm W2}=220.0^{+49.1}_{-74.1}$~days from the ICCF method. Using the $R_{\rm dust}-L_{\rm bol}$ relations derived from our low-mass AGN sample, the expected lags at $\log (L_{\rm bol}/\mathrm{erg\,s}^{-1})=43.67$ are only $\sim86$~days and $\sim110$~days for the W1 and W2 bands, respectively, about a factor of two smaller than the observed values. The discrepancy may partly arise because the adopted $L_{\rm bol}$ is based on an archival SDSS spectrum obtained approximately 17 years before the 2019 transient, and therefore may not represent the luminosity during the flare. The historical WISE light curves show a pronounced infrared flare, with the peak W1 and W2 flux densities reaching roughly 3--4 times the pre-flare levels. If the intrinsic nuclear luminosity during the flare had increased by a factor of $\sim4$, the expected W1- and W2-band lags would closely match the observed values. We also consider the infrared peak luminosities derived from the WISE light curves. Using $\log (L_{\rm W1,peak}/\mathrm{erg\,s}^{-1})=43.69$ and $\log (L_{\rm W2,peak}/\mathrm{erg\,s}^{-1})=43.60$ as luminosity proxies yields expected lags of $\sim140$~days and $\sim180$~days for the W1 and W2 bands, respectively. Although the peak infrared luminosities yield predictions closer to the observed lags, a modest difference still remains. Thus, the enhanced luminosity during the transient may partly account for the long dust lags, while the remaining difference may reflect the dust distribution and reverberation geometry associated with the transient outburst.

The dust response to a transient outburst may differ from that expected for a typical AGN dust structure. In particular, the intense radiation field of a TDE-like flare can sublimate dust in the innermost region, potentially moving the inner dust boundary outward and altering the subsequent infrared response. A similar scenario has been proposed for the TDE candidate PS16dtm, where the MIR echo was interpreted as emission from pre-existing circumnuclear dust. \citet{2017ApJ...850...63J} proposed that the intense radiation field during the flare evaporated dust at the inner edge, increasing the inner dust radius by a factor of $\sim7$. More recent modeling by \citet{2025ApJ...980L..17J} explicitly incorporated the dynamical evolution of the dust inner boundary during sublimation, finding that the inner dust radius increased from $\sim1.1$ to $\sim1.6$ pc during the outburst. Such transient-induced modification of the inner dust boundary may shift the responsivity-weighted dust radius outward, causing the observed infrared lags to exceed those expected from the $R_{\rm dust}-L_{\rm AGN}$ relation.

Another possibility is that the transient probes an extended circumnuclear dust distribution rather than the characteristic inner radius of a typical AGN torus. A statistical study of 23 optical TDEs suggests low dust covering factors at sub-parsec scales, with $f_c\sim0.01$ or less, indicating a relative paucity of compact circumnuclear dust around the central black hole \citep{2021ApJ...911...31J}. For the TDE candidate AT2019qiz, \citet{2025ApJ...988L..77W} modeled the IR echo and inferred an inner dust radius of $\gtrsim1.2$ pc, substantially larger than the expected dust sublimation radius for a system hosting a $\sim10^6\,M_\odot$ black hole. They interpreted this large radius as potentially reflecting a remnant of a previously extended AGN torus, in which the inner dust had disappeared while the outer dust structure persisted. Such an extended dust distribution may cause the transient IR response to be weighted toward larger radii, potentially contributing to the observed excess in the W1- and W2-band lags.

\section{Summary} \label{sec:Summary}
In this work, we perform a systematic dust RM study of low-mass AGNs selected from the catalog of \citet{2026HXG}, utilizing ZTF optical and WISE infrared light curves over a $\sim7$-year baseline. The main results are summarized as follows:

\begin{enumerate}
    \item Through a series of quality assessments, we obtain a core sample of 65 sources with both W1- and W2-band lag measurements. The dust lags obtained from the ICCF, JAVELIN, and PYROA methods are mutually consistent, with negligible systematic offsets ($<0.03$\,dex) and small scatters ($0.1$--$0.2$\,dex).

    \item The W2-band lags are systematically longer than the W1-band lags, with a median ratio of $\tau_{\rm W2}/\tau_{\rm W1}=1.24^{+0.33}_{-0.21}$, slightly higher than the lag ratios reported for higher-luminosity AGNs. The lag ratio also shows a weak anti-correlation with $L_{\rm bol}$, indicating a possible luminosity dependence of the dust temperature-stratified structure.

    \item For our low-mass AGN sample, we obtain multiple $R_{\rm dust}-L_{\rm AGN}$ relations, with slopes ranging from $\sim0.3$ to $\sim0.5$ across different lag-measurement methods and luminosity proxies. Combining our sample with the higher-luminosity AGN sample of \citet{2024ApJ...968...59M}, we establish unified $R_{\rm dust}-L_{\rm AGN}$ relations across nearly seven orders of magnitude in bolometric luminosity. The best-fit slopes are $0.47\pm0.01$ and $0.44\pm0.01$ for the W1 and W2 bands, respectively, with intrinsic scatters of $0.16\pm0.01$\,dex for both bands.

    \item The relatively shallow $R_{\rm dust}-L_{\rm AGN}$ slopes may be related to two effects. First, the luminosity-dependent SED variations revealed by the SPHEREx composite spectra suggest that individual luminosity proxies may not fully trace the intrinsic dust-heating luminosity. Second, the negative correlation between the lag deviations and Eddington ratio found in the combined sample may contribute to the mildly shallow $R_{\rm dust}-L_{\rm AGN}$ slopes, potentially reflecting the self-shadowing effect of slim disks.

    \item We find a low-mass AGN associated with the TDE-like transient AT2019dqv. The observed W1 and W2 dust lags are substantially longer than expected from the $R_{\rm dust}-L_{\rm AGN}$ relations. Accounting for the enhanced luminosity during the transient brings the expected lags closer to the observed values, but does not fully resolve the discrepancy, which may be related to the dust distribution and reverberation geometry associated with the transient outburst.
\end{enumerate}

Overall, our results extend dust RM studies to low-mass AGNs and establish unified $R_{\rm dust}-L_{\rm AGN}$ relations across a broad luminosity range. However, probing lower-luminosity AGNs requires measurements of shorter dust lags, which remain challenging with current infrared monitoring data. Future joint optical and near-infrared monitoring with the Vera C. Rubin Observatory's Legacy Survey of Space and Time and the Nancy Grace Roman Space Telescope will offer promising opportunities to address these challenges and extend dust RM studies to lower-luminosity AGNs hosting lower-mass black holes.

\section*{ACKNOWLEDGMENTS}

We acknowledge support from the National Key R\&D Program of China (2025YFA1614100, 2025YFA1614101) and the National Natural Science Foundation of China (No. 12133001). Hengxiao Guo acknowledges support from the National Natural Science Foundation of China (Nos. 12473018 and 12522304) and the Overseas Center Platform Projects of the Chinese Academy of Sciences (No. 178GJHZ2023184MI).

This work makes use of data products from WISE, which is a joint project of the University of California, Los Angeles, and the Jet Propulsion Laboratory/California Institute of Technology, funded by the National Aeronautics and Space Administration. This work also makes use of data products from NEOWISE, which is a joint project of the Jet Propulsion Laboratory/California Institute of Technology and the University of California, Los Angeles. NEOWISE is funded by the National Aeronautics and Space Administration.

This work is based on observations obtained with the Samuel Oschin Telescope 48 inch and the 60 inch Telescope at the Palomar Observatory as part of the Zwicky Transient Facility project. ZTF is supported by the National Science Foundation under grants No. AST-1440341 and AST-2034437 and a collaboration including current partners Caltech, IPAC, the Oskar Klein Center at Stockholm University, the University of Maryland, University of California, Berkeley, the University of Wisconsin at Milwaukee, University of Warwick, Ruhr University Bochum, Cornell University, Northwestern University, and Drexel University. Operations are conducted by COO, IPAC, and UW.

This publication makes use of data products from the Spectro-Photometer for the History of the Universe, Epoch of Reionization and Ices Explorer (SPHEREx), which is a joint project of the Jet Propulsion Laboratory and the California Institute of Technology, and is funded by the National Aeronautics and Space Administration.


\bibliography{sample701}{}
\bibliographystyle{aasjournalv7}

\appendix

This appendix presents supplementary figures and regression results supporting the lag measurements and $R_{\rm dust}-L_{\rm AGN}$ relations discussed in the main text. Figure~\ref{fig:Lag_results} shows the light curves and detailed lag-measurement results for six additional representative sources spanning a range of bolometric luminosities and black hole masses within our core sample. Figure~\ref{fig:lag_bin_comparison} compares the W1- and W2-band lag measurements obtained under four different optical binning schemes (3-day, 5-day, 10-day, and 15-day intervals) for two representative sources, J122342.81+581446.1 and J135724.51+652505.9. Figure~\ref{fig:Lag_comparison_methods} compares the ADC-corrected W1- and W2-band lags obtained with ICCF, JAVELIN, and PYROA. Figure~\ref{fig:R-Lag_ratio_dependence} shows the dependence of the W2-to-W1 lag ratio on AGN bolometric luminosity and Eddington ratio. Figure~\ref{fig:R-L_relations} presents the $R_{\rm dust}-L_{\rm AGN}$ relations based on alternative luminosity proxies, including $L_{5100}$, $L_{5100,\mathrm{AGN}}$, $L_{\rm W1}$, and $L_{\rm W2}$, with the corresponding regression parameters listed in Table~\ref{tab:R-L Relations}.

\renewcommand{\thetable}{A\arabic{table}}
\setcounter{table}{0}
\begin{sidewaystable*}
    \begin{center}
    \caption{The Best-fitting Regression Parameters of the $R_{\rm dust}-L_{\rm AGN}$ Relations.}
    \vspace{-1em}
    \renewcommand{\arraystretch}{0.93}
    \label{tab:R-L Relations}
    \setlength\tabcolsep{1.5mm}{
    \begin{tabular}{cccccccccccc}
        \toprule
        \toprule
        \multirow{2}{*}{Size} & \multirow{2}{*}{Luminosity}  & \multicolumn{5}{c}{Without ADC} & \multicolumn{5}{c}{With ADC} \\
        &  & $\alpha_{\mathrm{free}}$ & $\beta_{\mathrm{free}}$ & $\sigma_{\mathrm{free}}$ & $\beta_{\mathrm{fix}}$ & $\sigma_{\mathrm{fix}}$ & $\alpha_{\mathrm{free}}$ & $\beta_{\mathrm{free}}$ & $\sigma_{\mathrm{free}}$ & $\beta_{\mathrm{fix}}$ & $\sigma_{\mathrm{fix}}$ \\
        (1) & (2) & (3) & (4) & (5) & (6) & (7) & (8) & (9) & (10) & (11) & (12) \\
        \hline

        $R_{\text{W1,ICCF}}$ & $L_{\mathrm{bol}}$ & $0.43 \pm 0.07$ & $1.61 \pm 0.06$ & $0.15 \pm 0.03$ & $1.57 \pm 0.03$ & $0.15 \pm 0.03$ & $0.43 \pm 0.07$ & $1.65 \pm 0.05$& $0.13 \pm 0.03$ & $1.60 \pm 0.03$ & $0.14 \pm 0.03$ \\
        $R_{\text{W1,ICCF,Merged}}$ & $L_{\mathrm{bol,Merged}}$ & $0.46 \pm 0.01$ & $2.01 \pm 0.02$ & $0.18 \pm 0.01$ & $1.95 \pm 0.01$ & $0.18 \pm 0.01$ & $0.47 \pm 0.01$ & $2.05 \pm 0.02$ & $0.16 \pm 0.01$ & $2.00 \pm 0.01$ & $0.16 \pm 0.01$ \\
        $R_{\text{W1,JAVELIN}}$ & $L_{\mathrm{bol}}$ & $0.49 \pm 0.07$ & $1.53 \pm 0.06$ & $0.22 \pm 0.03$ & $1.53 \pm 0.04$ & $0.21 \pm 0.03$ & $0.46 \pm 0.07$ & $1.62 \pm 0.06$& $0.19 \pm 0.03$ & $1.59 \pm 0.03$ & $0.19 \pm 0.03$ \\
        $R_{\text{W1,PYROA}}$ & $L_{\mathrm{bol}}$ & $0.41 \pm 0.07$ & $1.61 \pm 0.05$ & $0.20 \pm 0.03$ & $1.55 \pm 0.03$ & $0.21 \pm 0.03$ & $0.45 \pm 0.06$ & $1.63 \pm 0.04$ & $0.18 \pm 0.02$& $1.60 \pm 0.03$ & $0.18 \pm 0.02$ \\
        \hline
        $R_{\text{W1,ICCF}}$ & $L_{5100}$ & $0.39 \pm 0.07$ & $1.88 \pm 0.04$ & $0.14 \pm 0.04$ & $1.88 \pm 0.04$ & $0.15 \pm 0.04$ & $0.40 \pm 0.06$ & $1.91 \pm 0.03$ & $0.12 \pm 0.04$ & $1.91 \pm 0.03$ & $0.13 \pm 0.04$ \\
        $R_{\text{W1,JAVELIN}}$ & $L_{5100}$ & $0.43 \pm 0.06$ & $1.85 \pm 0.04$ & $0.22 \pm 0.03$ & $1.85 \pm 0.04$ & $0.22 \pm 0.03$ & $0.43 \pm 0.07$ & $1.90 \pm 0.03$ & $0.20 \pm 0.03$ & $1.90 \pm 0.03$ & $0.20 \pm 0.03$ \\
        $R_{\text{W1,PYROA}}$ & $L_{5100}$ & $0.39 \pm 0.06$ & $1.86 \pm 0.03$ & $0.20 \pm 0.03$ & $1.85 \pm 0.03$ & $0.21 \pm 0.03$ & $0.41 \pm 0.04$ & $1.91 \pm 0.03$ & $0.16 \pm 0.02$ & $1.91 \pm 0.03$ & $0.17 \pm 0.02$ \\
        \hline
        $R_{\text{W1,ICCF}}$ & $L_{5100,\mathrm{AGN}}$ & $0.31 \pm 0.06$ & $2.09 \pm 0.05$ & $0.17 \pm 0.04$ & $2.20 \pm 0.04$ & $0.20 \pm 0.03$ & $0.31 \pm 0.06$ & $2.12 \pm 0.05$ & $0.15 \pm 0.03$ & $2.23 \pm 0.04$ & $0.19 \pm 0.03$ \\
        $R_{\text{W1,JAVELIN}}$ & $L_{5100,\mathrm{AGN}}$ & $0.36 \pm 0.06$ & $2.08 \pm 0.06$ & $0.24 \pm 0.03$ & $2.17 \pm 0.04$ & $0.26 \pm 0.03$ & $0.34 \pm 0.06$ & $2.13 \pm 0.05$ & $0.22 \pm 0.03$ & $2.23 \pm 0.04$ & $0.23 \pm 0.03$ \\
        $R_{\text{W1,PYROA}}$ & $L_{5100,\mathrm{AGN}}$ & $0.29 \pm 0.06$ & $2.06 \pm 0.05$ & $0.23 \pm 0.03$ & $2.19 \pm 0.04$ & $0.26 \pm 0.03$ & $0.33 \pm 0.05$ & $2.13 \pm 0.04$ & $0.20 \pm 0.02$ & $2.24 \pm 0.03$ & $0.23 \pm 0.03$ \\
        \hline
        $R_{\text{W1,ICCF}}$ & $L_{\rm W1}$ & $0.36 \pm 0.05$ & $1.87 \pm 0.03$ & $0.11 \pm 0.04$ & $1.85 \pm 0.03$ & $0.13 \pm 0.04$ & $0.35 \pm 0.05$ & $1.91 \pm 0.03$ & $0.09 \pm 0.04$ & $1.89 \pm 0.03$ & $0.12 \pm 0.04$ \\
        $R_{\text{W1,JAVELIN}}$ & $L_{\rm W1}$ & $0.40 \pm 0.05$ & $1.82 \pm 0.04$ & $0.21 \pm 0.03$ & $1.82 \pm 0.04$ & $0.22 \pm 0.03$ & $0.36 \pm 0.06$ & $1.90 \pm 0.03$ & $0.19 \pm 0.03$ & $1.89 \pm 0.03$ & $0.20 \pm 0.03$ \\
        $R_{\text{W1,PYROA}}$ & $L_{\rm W1}$ & $0.33 \pm 0.05$ & $1.85 \pm 0.03$ & $0.19 \pm 0.03$ & $1.84 \pm 0.03$ & $0.21 \pm 0.03$ & $0.35 \pm 0.04$ & $1.91 \pm 0.03$ & $0.17 \pm 0.02$ & $1.91 \pm 0.03$ & $0.19 \pm 0.03$ \\
        \hline
        $R_{\text{W1,ICCF}}$ & $L_{\rm W2}$ & $0.35 \pm 0.05$ & $1.90 \pm 0.03$ & $0.11 \pm 0.04$ & $1.90 \pm 0.03$ & $0.13 \pm 0.04$ & $0.34 \pm 0.05$ & $1.94 \pm 0.03$ & $0.09 \pm 0.04$ & $1.94 \pm 0.03$ & $0.12 \pm 0.03$ \\
        $R_{\text{W1,JAVELIN}}$ & $L_{\rm W2}$ & $0.39 \pm 0.05$ & $1.87 \pm 0.04$ & $0.21 \pm 0.03$ & $1.87 \pm 0.04$ & $0.21 \pm 0.03$ & $0.35 \pm 0.05$ & $1.94 \pm 0.03$ & $0.19 \pm 0.03$ & $1.95 \pm 0.03$ & $0.20 \pm 0.03$ \\
        $R_{\text{W1,PYROA}}$ & $L_{\rm W2}$ & $0.32 \pm 0.05$ & $1.89 \pm 0.03$ & $0.19 \pm 0.03$ & $1.90 \pm 0.03$ & $0.21 \pm 0.03$ & $0.34 \pm 0.04$ & $1.95 \pm 0.03$ & $0.18 \pm 0.02$ & $1.96 \pm 0.03$ & $0.20 \pm 0.03$ \\
        \hline
        $R_{\text{W2,ICCF}}$ & $L_{\mathrm{bol}}$ & $0.47 \pm 0.07$ & $1.71 \pm 0.05$ & $0.14 \pm 0.03$ & $1.69 \pm 0.03$ & $0.13 \pm 0.03$ & $0.46 \pm 0.06$& $1.73 \pm 0.05$ & $0.13 \pm 0.03$ & $1.70 \pm 0.03$ & $0.13 \pm 0.03$ \\
        $R_{\text{W2,ICCF,Merged}}$ & $L_{\mathrm{bol,Merged}}$ & $0.44 \pm 0.01$ & $2.16 \pm 0.02$ & $0.17 \pm 0.01$ & $2.06 \pm 0.01$ & $0.18 \pm 0.01$ & $0.44 \pm 0.01$ & $2.19 \pm 0.02$ & $0.16 \pm 0.01$ & $2.09 \pm 0.01$ & $0.16 \pm 0.01$ \\
        $R_{\text{W2,JAVELIN}}$ & $L_{\mathrm{bol}}$ & $0.47 \pm 0.07$ & $1.70 \pm 0.05$ & $0.17 \pm 0.03$ & $1.69 \pm 0.03$ & $0.17 \pm 0.03$ & $0.46 \pm 0.06$ & $1.72 \pm 0.05$ & $0.16 \pm 0.03$ & $1.70 \pm 0.03$ & $0.16 \pm 0.03$ \\
        $R_{\text{W2,PYROA}}$ & $L_{\mathrm{bol}}$ & $0.42 \pm 0.07$ & $1.72 \pm 0.05$ & $0.20 \pm 0.03$ & $1.67 \pm 0.03$ & $0.21 \pm 0.03$ & $0.46 \pm 0.05$ & $1.71 \pm 0.04$ & $0.18 \pm 0.02$ & $1.69 \pm 0.03$ & $0.18 \pm 0.02$ \\
        \hline
        $R_{\text{W2,ICCF}}$ & $L_{5100}$ & $0.43 \pm 0.06$ & $2.01 \pm 0.03$ & $0.13 \pm 0.03$ & $2.01 \pm 0.03$ & $0.13 \pm 0.03$ & $0.43 \pm 0.06$ & $2.03 \pm 0.03$ & $0.12 \pm 0.03$ & $2.02 \pm 0.03$ & $0.12 \pm 0.03$ \\
        $R_{\text{W2,JAVELIN}}$ & $L_{5100}$ & $0.44 \pm 0.06$ & $2.00 \pm 0.04$ & $0.17 \pm 0.03$ & $2.00 \pm 0.04$ & $0.17 \pm 0.03$ & $0.43 \pm 0.06$ & $2.01 \pm 0.03$ & $0.16 \pm 0.03$ & $2.00 \pm 0.03$ & $0.16 \pm 0.03$ \\
        $R_{\text{W2,PYROA}}$ & $L_{5100}$ & $0.40 \pm 0.06$ & $1.98 \pm 0.03$ & $0.20 \pm 0.03$ & $1.98 \pm 0.03$ & $0.20 \pm 0.03$ & $0.43 \pm 0.04$ & $2.00 \pm 0.03$ & $0.15 \pm 0.02$ & $2.01 \pm 0.03$ & $0.15 \pm 0.02$ \\
        \hline
        $R_{\text{W2,ICCF}}$ & $L_{5100,\mathrm{AGN}}$ & $0.33 \pm 0.06$ & $2.22 \pm 0.05$ & $0.16 \pm 0.03$ & $2.33 \pm 0.04$ & $0.19 \pm 0.03$ & $0.33 \pm 0.05$ & $2.24 \pm 0.05$ & $0.16 \pm 0.03$ & $2.34 \pm 0.03$ & $0.18 \pm 0.03$ \\
        $R_{\text{W2,JAVELIN}}$ & $L_{5100,\mathrm{AGN}}$ & $0.34 \pm 0.06$ & $2.22 \pm 0.06$ & $0.20 \pm 0.03$ & $2.33 \pm 0.04$ & $0.22 \pm 0.03$ & $0.34 \pm 0.05$ & $2.23 \pm 0.05$ & $0.19 \pm 0.03$ & $2.34 \pm 0.04$ & $0.21 \pm 0.03$ \\
        $R_{\text{W2,PYROA}}$ & $L_{5100,\mathrm{AGN}}$ & $0.30 \pm 0.06$ & $2.18 \pm 0.05$ & $0.22 \pm 0.03$ & $2.31 \pm 0.04$ & $0.25 \pm 0.03$ & $0.35 \pm 0.04$ & $2.23 \pm 0.04$ & $0.19 \pm 0.02$ & $2.33 \pm 0.03$ & $0.21 \pm 0.03$ \\
        \hline
        $R_{\text{W2,ICCF}}$ & $L_{\rm W1}$ & $0.38 \pm 0.05$ & $1.98 \pm 0.03$ & $0.11 \pm 0.03$ & $1.97 \pm 0.03$ & $0.12 \pm 0.03$ & $0.37 \pm 0.05$ & $2.01 \pm 0.03$ & $0.10 \pm 0.03$ & $2.00 \pm 0.03$ & $0.12 \pm 0.03$ \\
        $R_{\text{W2,JAVELIN}}$ & $L_{\rm W1}$ & $0.38 \pm 0.05$ & $1.98 \pm 0.03$ & $0.14 \pm 0.03$ & $1.97 \pm 0.03$ & $0.16 \pm 0.03$ & $0.38 \pm 0.05$ & $1.99 \pm 0.03$ & $0.13 \pm 0.03$ & $1.99 \pm 0.03$ & $0.14 \pm 0.03$ \\
        $R_{\text{W2,PYROA}}$ & $L_{\rm W1}$ & $0.35 \pm 0.05$ & $1.97 \pm 0.03$ & $0.19 \pm 0.03$ & $1.96 \pm 0.03$ & $0.20 \pm 0.03$ & $0.38 \pm 0.04$ & $2.00 \pm 0.03$ & $0.15 \pm 0.02$ & $2.00 \pm 0.03$ & $0.17 \pm 0.02$ \\
        \hline
        $R_{\text{W2,ICCF}}$ & $L_{\rm W2}$ & $0.37 \pm 0.05$ & $2.03 \pm 0.03$ & $0.10 \pm 0.03$ & $2.03 \pm 0.03$ & $0.12 \pm 0.03$ & $0.36 \pm 0.04$ & $2.05 \pm 0.03$ & $0.10 \pm 0.03$ & $2.05 \pm 0.03$ & $0.12 \pm 0.03$ \\
        $R_{\text{W2,JAVELIN}}$ & $L_{\rm W2}$ & $0.36 \pm 0.05$ & $2.02 \pm 0.03$ & $0.14 \pm 0.03$ & $2.03 \pm 0.03$ & $0.16 \pm 0.03$ & $0.36 \pm 0.04$ & $2.03 \pm 0.03$ & $0.13 \pm 0.03$ & $2.04 \pm 0.03$ & $0.15 \pm 0.03$ \\
        $R_{\text{W2,PYROA}}$ & $L_{\rm W2}$ & $0.34 \pm 0.05$ & $2.01 \pm 0.03$ & $0.19 \pm 0.03$ & $2.02 \pm 0.03$ & $0.21 \pm 0.03$ & $0.37 \pm 0.04$ & $2.04 \pm 0.03$ & $0.16 \pm 0.02$ & $2.05 \pm 0.03$ & $0.17 \pm 0.03$ \\

        \bottomrule
    \end{tabular}}
    \end{center}
    \vspace{-0.5em}
    \textbf{Note.} Column (1): Torus size used in the fitting. 
Column (2): Luminosity used in the fitting. 
Columns (3)–(5): Slope ($\alpha_{\mathrm{free}}$), intercept ($\beta_{\mathrm{free}}$), and intrinsic scatter ($\sigma_{\mathrm{free}}$) of the best fits with $\alpha$ treated as a free parameter (without ADC). 
Columns (6)–(7): Intercept ($\beta_{\mathrm{fix}}$) and intrinsic scatter ($\sigma_{\mathrm{fix}}$) of the best fits with $\alpha$ fixed at 0.5 (without ADC). 
Columns (8)–(10): Slope ($\alpha_{\mathrm{free}}$), intercept ($\beta_{\mathrm{free}}$), and intrinsic scatter ($\sigma_{\mathrm{free}}$) of the best fits with $\alpha$ treated as a free parameter (with ADC). 
Columns (11)–(12): Intercept ($\beta_{\mathrm{fix}}$) and intrinsic scatter ($\sigma_{\mathrm{fix}}$) of the best fits with $\alpha$ fixed at 0.5 (with ADC).
\end{sidewaystable*}

\renewcommand{\thefigure}{A\arabic{figure}}
\setcounter{figure}{0}

\begin{figure*}[t]
    \centering
    \begin{minipage}{0.45\textwidth}
        \includegraphics[width=\linewidth]{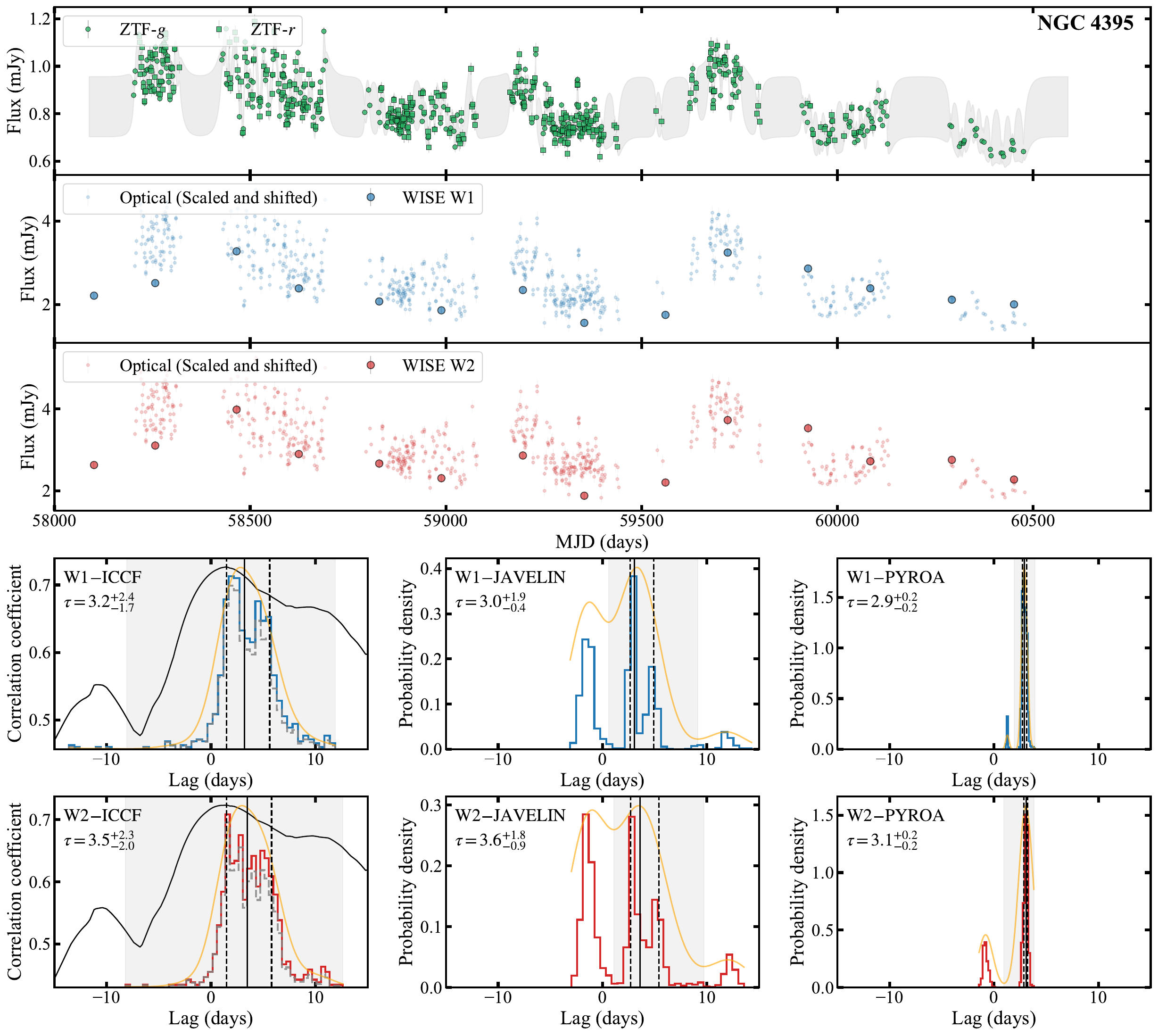}
    \end{minipage}%
    \begin{minipage}{0.45\textwidth}
        \includegraphics[width=\linewidth]{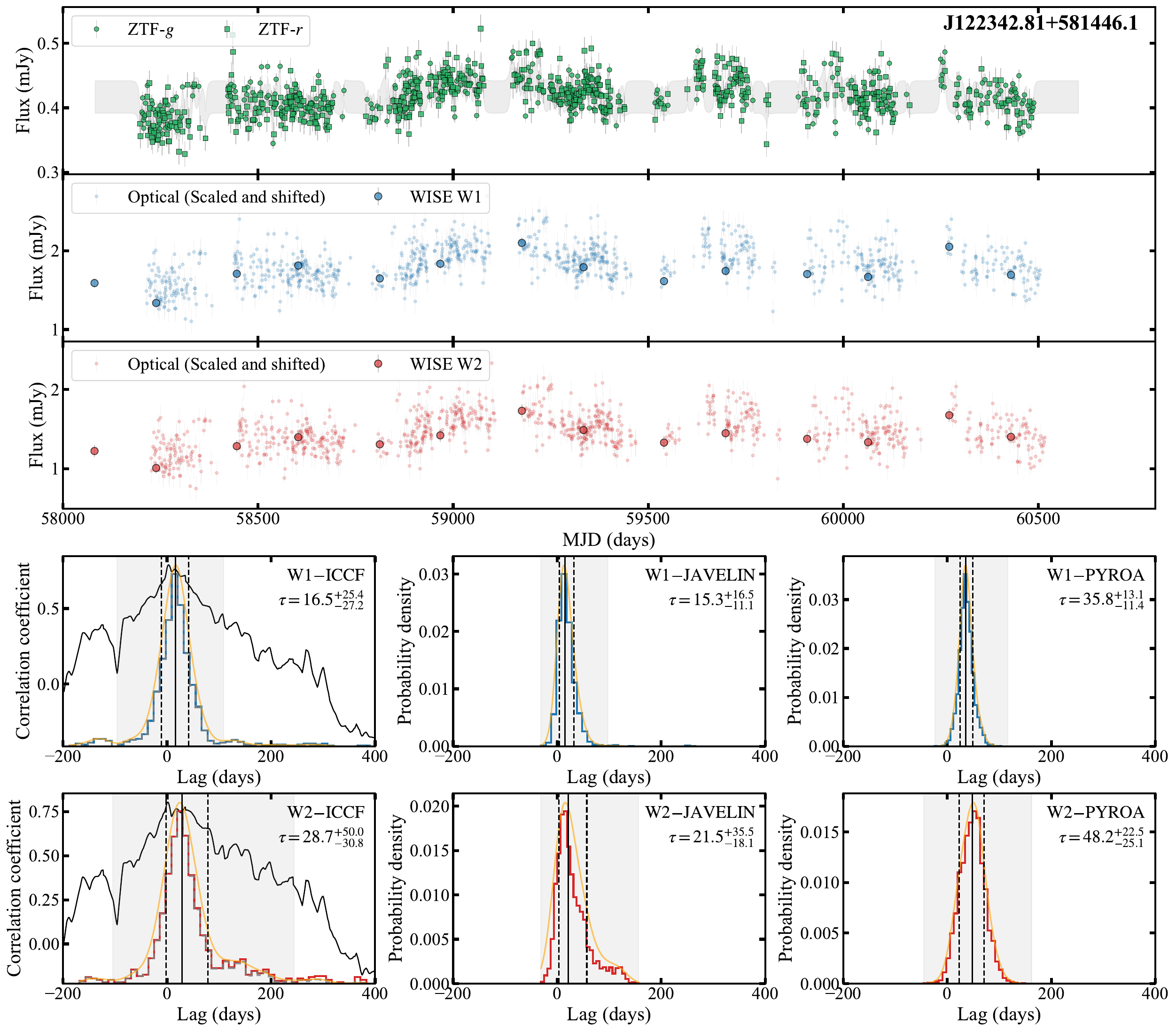}
    \end{minipage}

    \begin{minipage}{0.45\textwidth}
        \includegraphics[width=\linewidth]{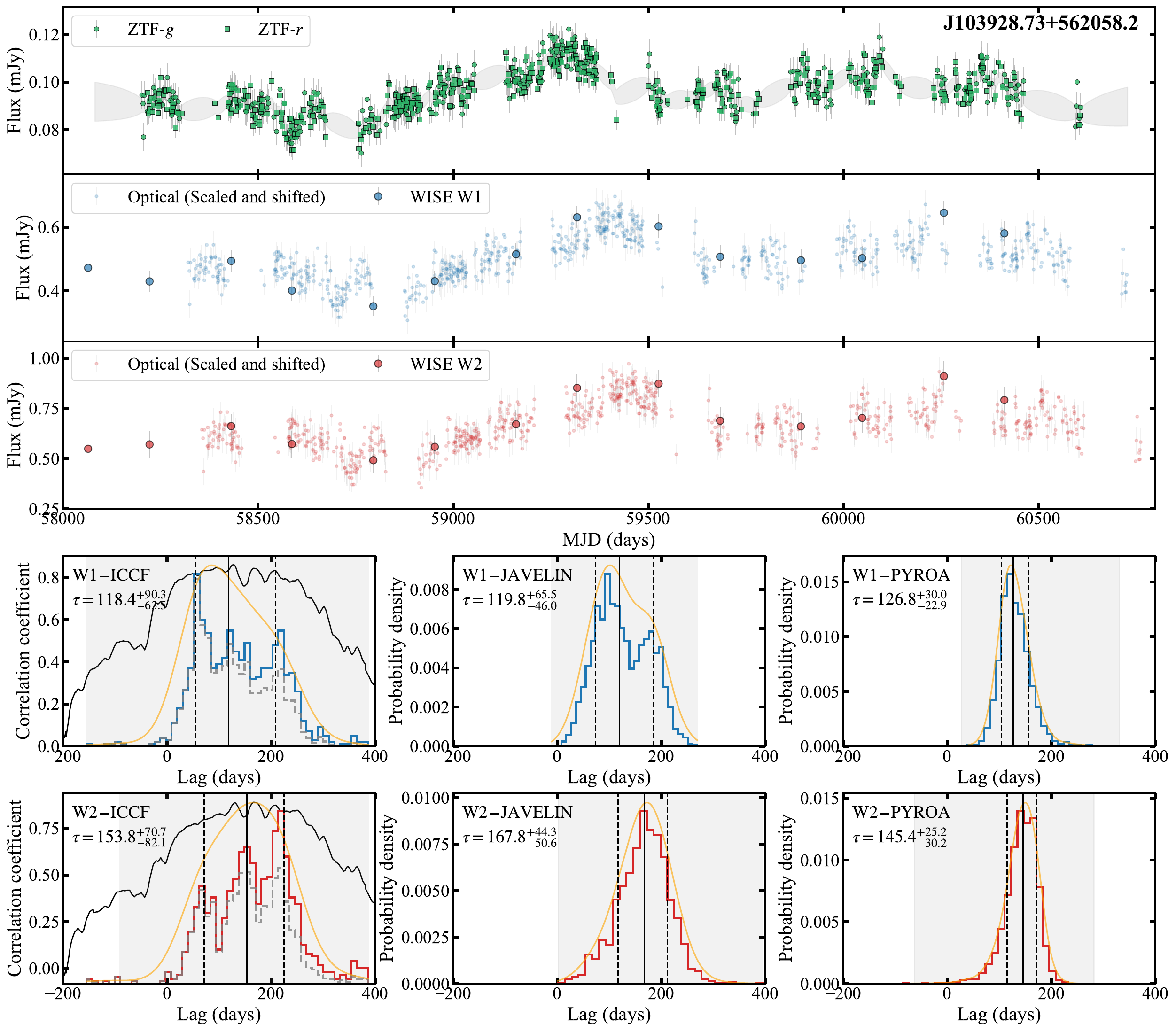}
    \end{minipage}%
    \begin{minipage}{0.45\textwidth}
        \includegraphics[width=\linewidth]{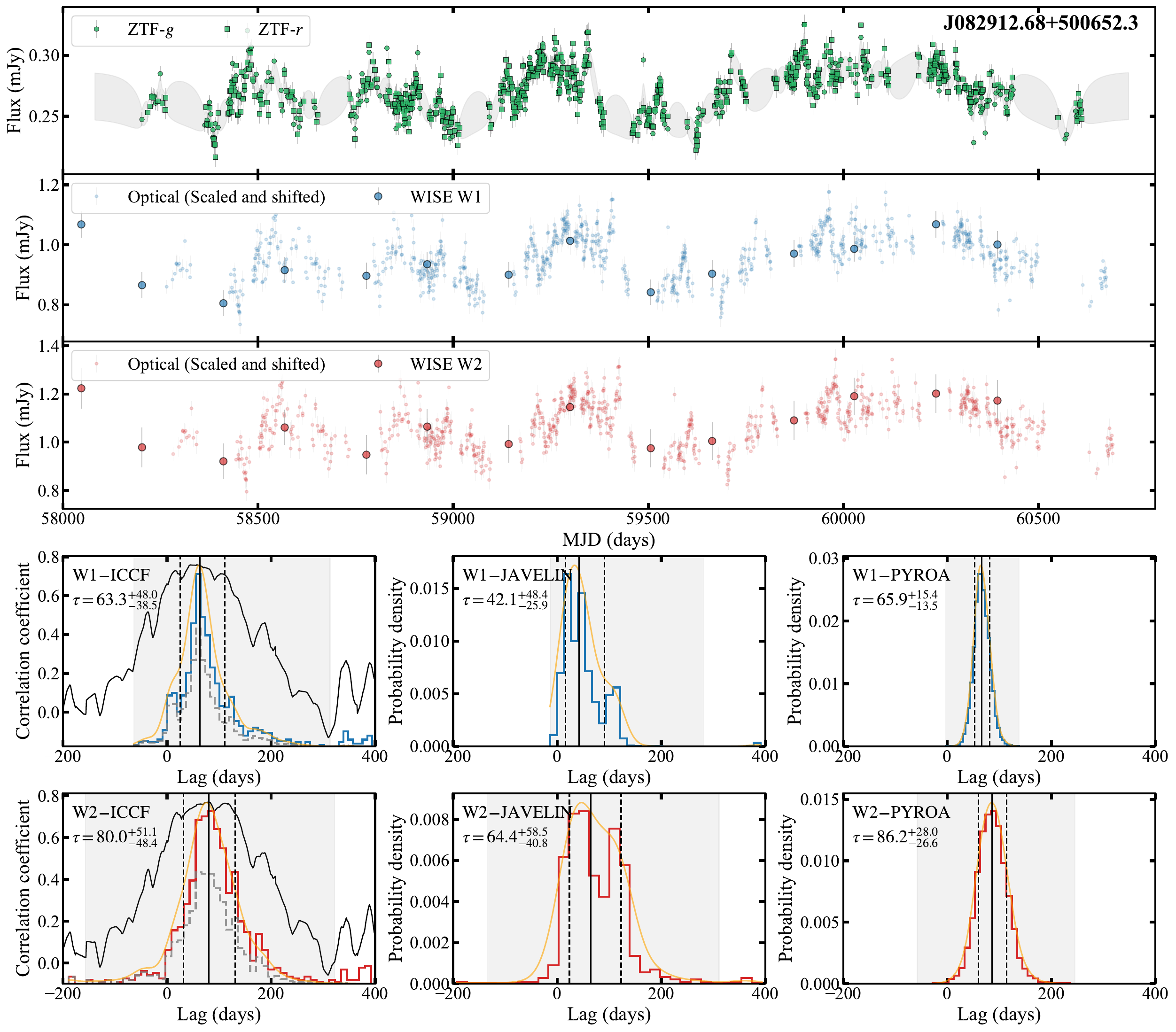}
    \end{minipage}

    \begin{minipage}{0.45\textwidth}
        \includegraphics[width=\linewidth]{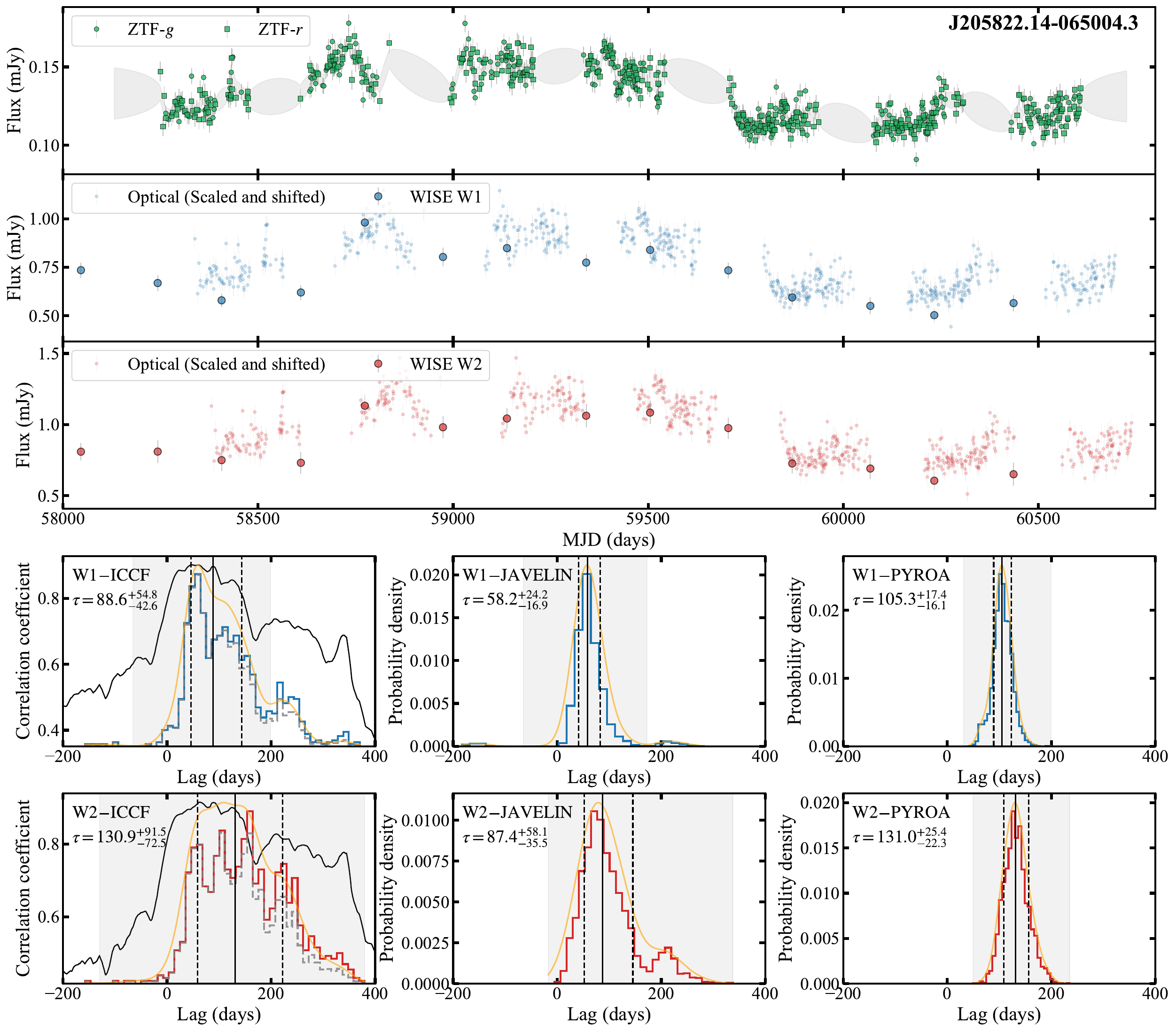}
    \end{minipage}%
    \begin{minipage}{0.45\textwidth}
        \includegraphics[width=\linewidth]{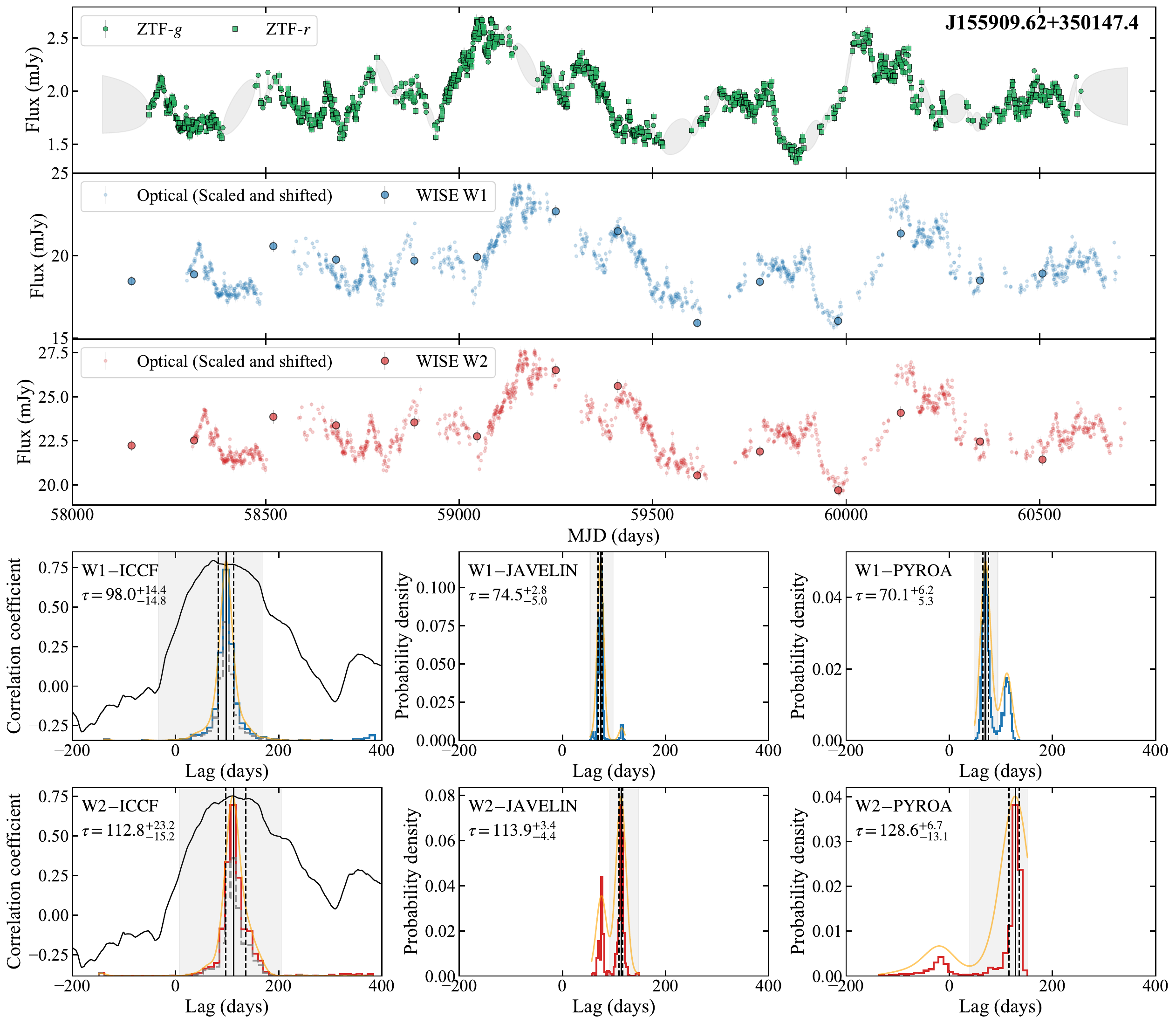}
    \end{minipage}
    \caption{Same as Figure~\ref{fig:result_35} but for six additional sources, including J082912.68+500652.3, J103928.73+562058.2, J122342.81+581446.1, NGC~4395, J155909.62+350147.4, and J205822.14-065004.3.}
    \label{fig:Lag_results}
\end{figure*}

\begin{figure*}[h]
    \centering

    \begin{subfigure}{0.99\textwidth}
        \centering
        \includegraphics[width=\linewidth]{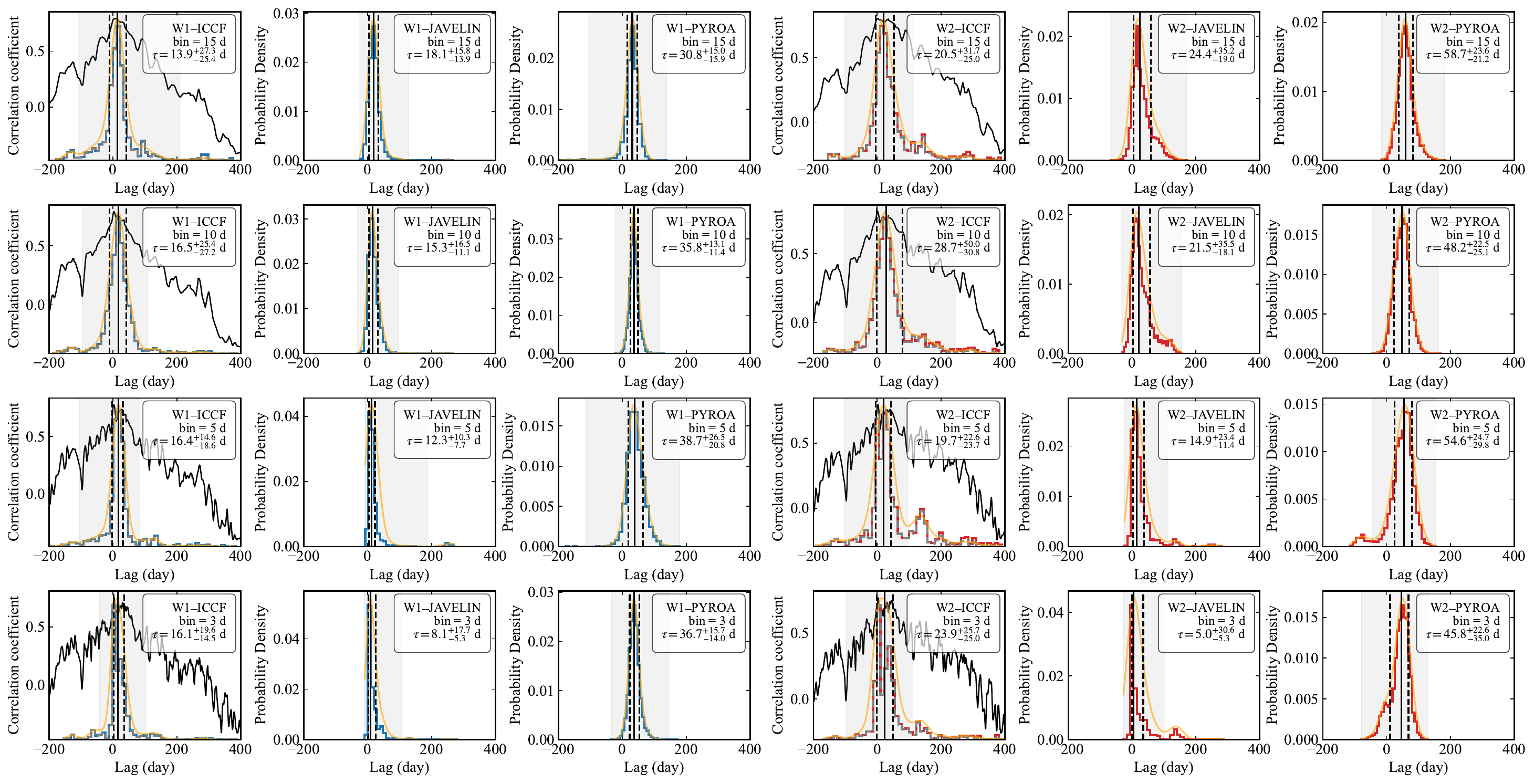}
        \caption{J122342.81+581446.1}
    \end{subfigure}

    \begin{subfigure}{0.99\textwidth}
        \centering
        \includegraphics[width=\linewidth]{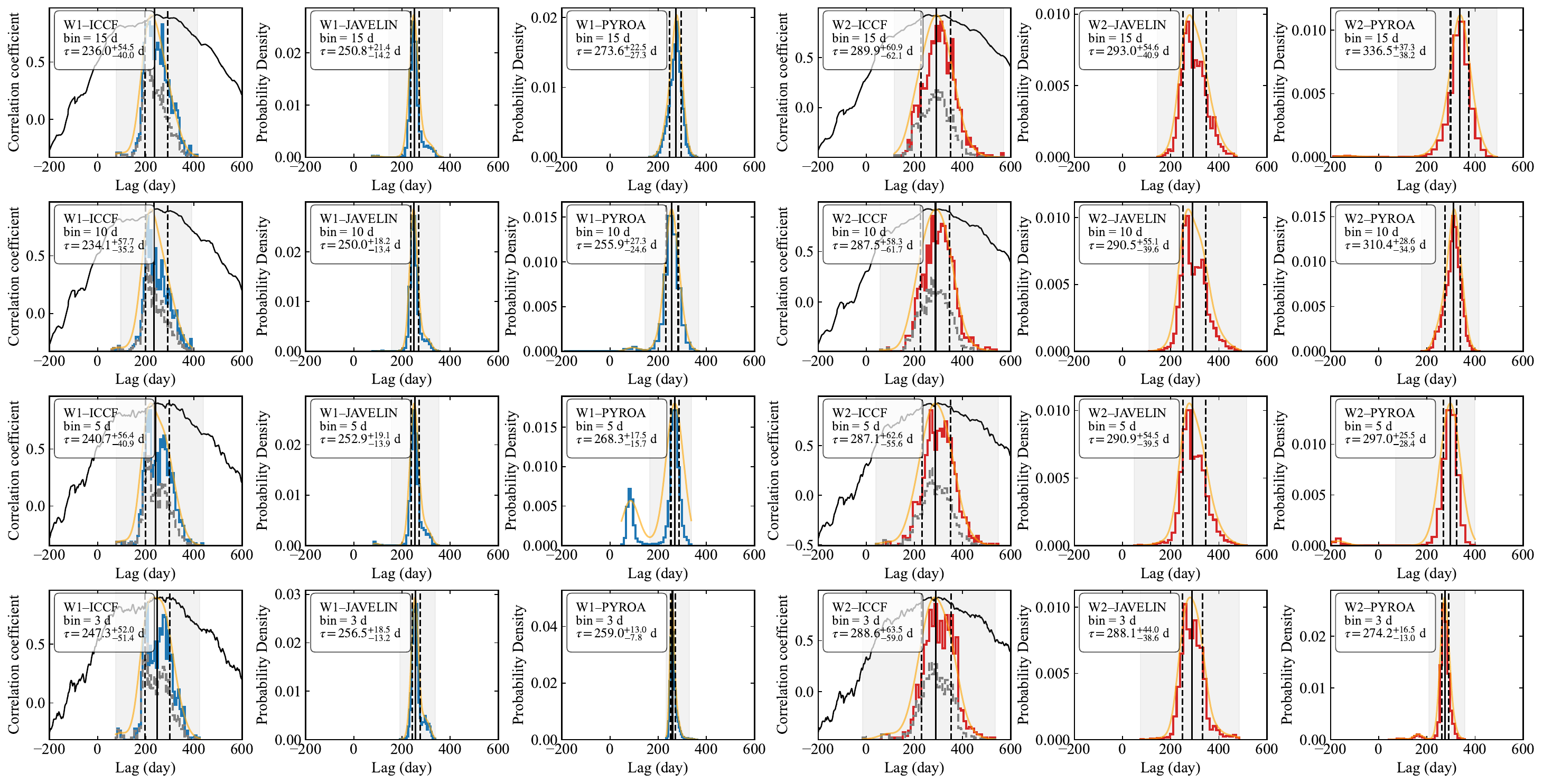}
        \caption{J135724.51+652505.9}
    \end{subfigure}

    \caption{
    Comparison of lag measurements based on optical light curves with different binning schemes. Panels (a) and (b) show the results for J122342.81+581446.1 and J135724.51+652505.9, respectively. Each source is presented in a $4 \times 6$ grid of subpanels, with the four rows corresponding to 15-day, 10-day, 5-day, and 3-day binned optical light curves (from top to bottom). Columns 1--3 show the W1-band lag measurements obtained using ICCF, JAVELIN, and PYROA, respectively, while columns 4--6 present the corresponding W2-band results.
    }

    \label{fig:lag_bin_comparison}
\end{figure*}

\begin{figure*}[]
    \centering
    \includegraphics[width=0.90\textwidth]{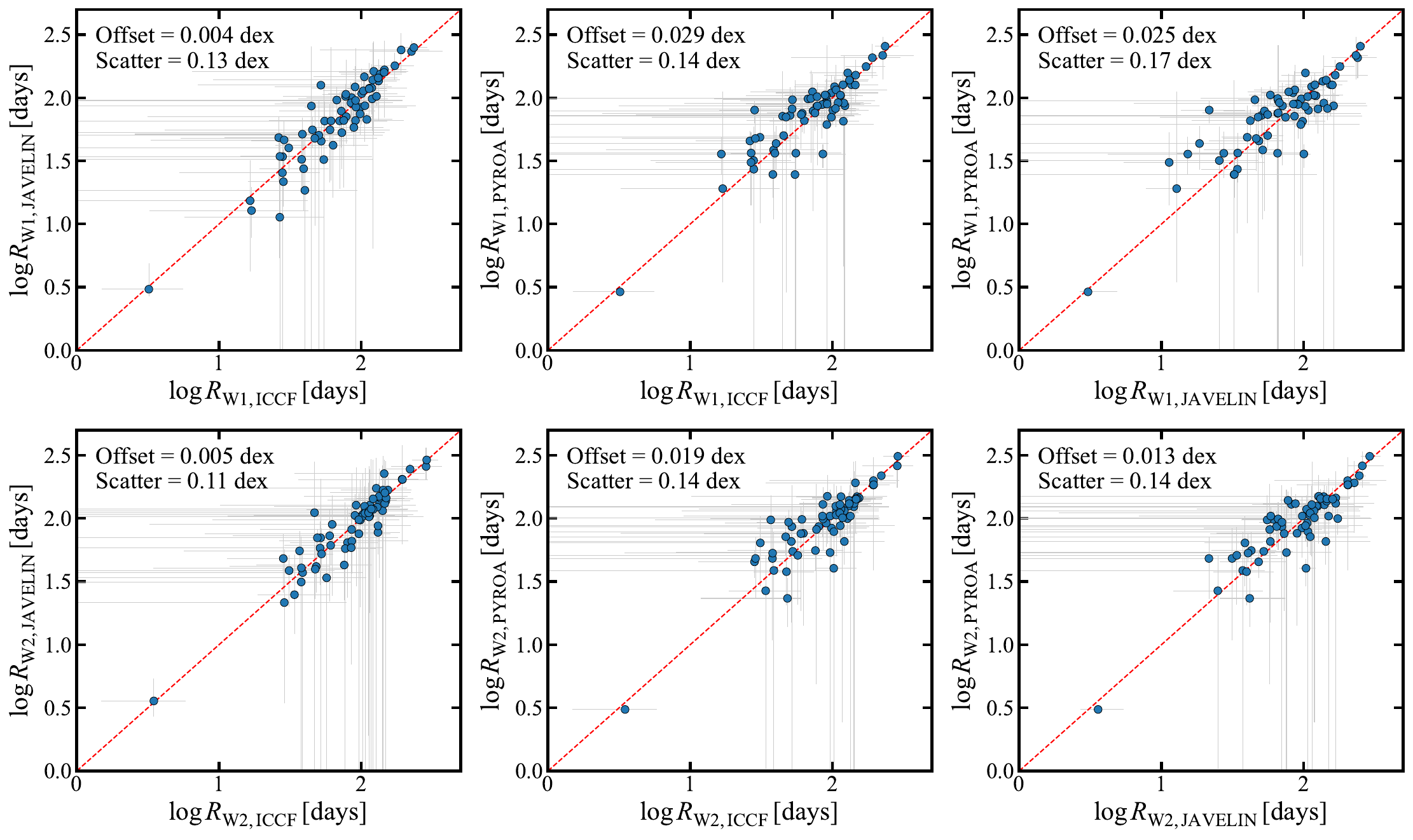} 
    \caption{Comparison of the ADC-corrected W1- and W2-band lags obtained from different methods. The top and bottom rows show the results for the W1 and W2 bands, respectively. The three columns compare ICCF versus JAVELIN lags (left), ICCF versus PYROA lags (middle), and JAVELIN versus PYROA lags (right). The red dashed line represents the 1:1 relation.}
    \label{fig:Lag_comparison_methods}
\end{figure*}

\begin{figure*}[t]
    \centering
    \includegraphics[width=0.90\textwidth]{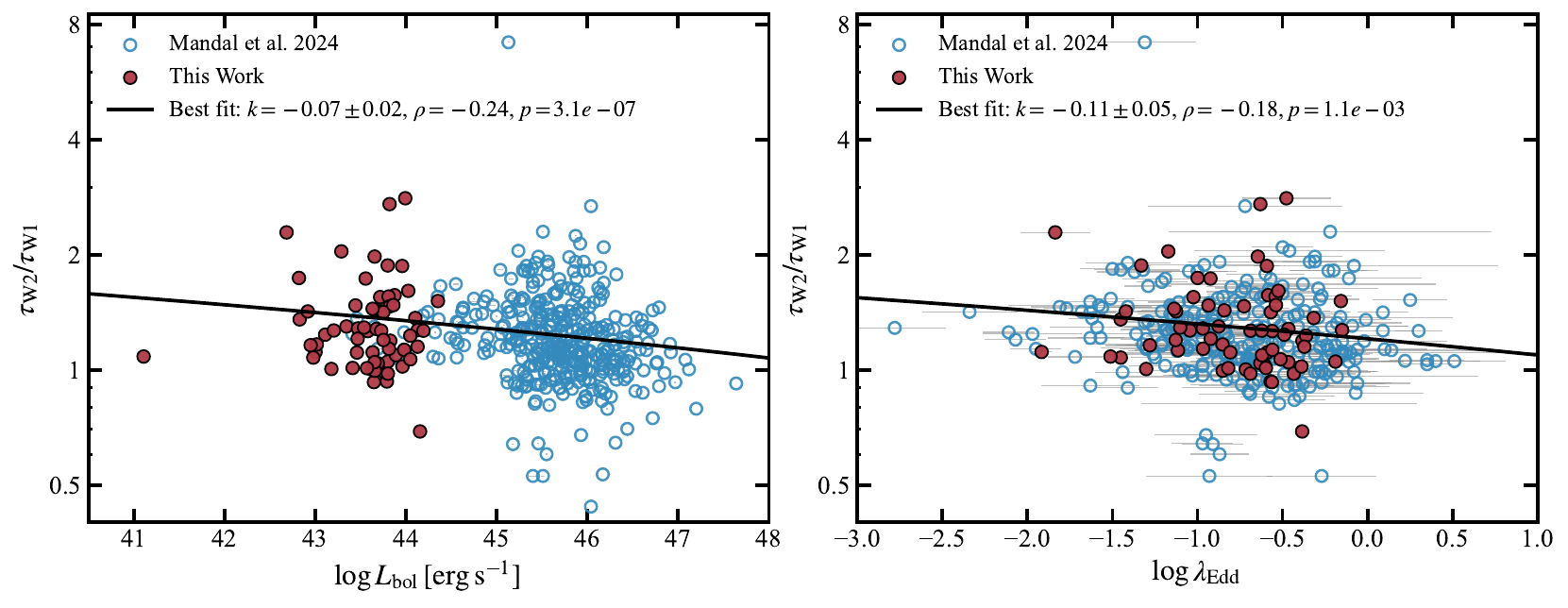} 
    \caption{Dependence of the W2-to-W1 lag ratio on AGN bolometric luminosity (left) and Eddington ratio (right). The red filled circles represent the measurements from this work, while the blue open circles show those from \citet{2024ApJ...968...59M}. The black lines denote the best-fit relations.}
    \label{fig:R-Lag_ratio_dependence}
\end{figure*}

\begin{figure*}[h]
    \centering
    \begin{minipage}{0.5\textwidth}
        \includegraphics[width=\linewidth]{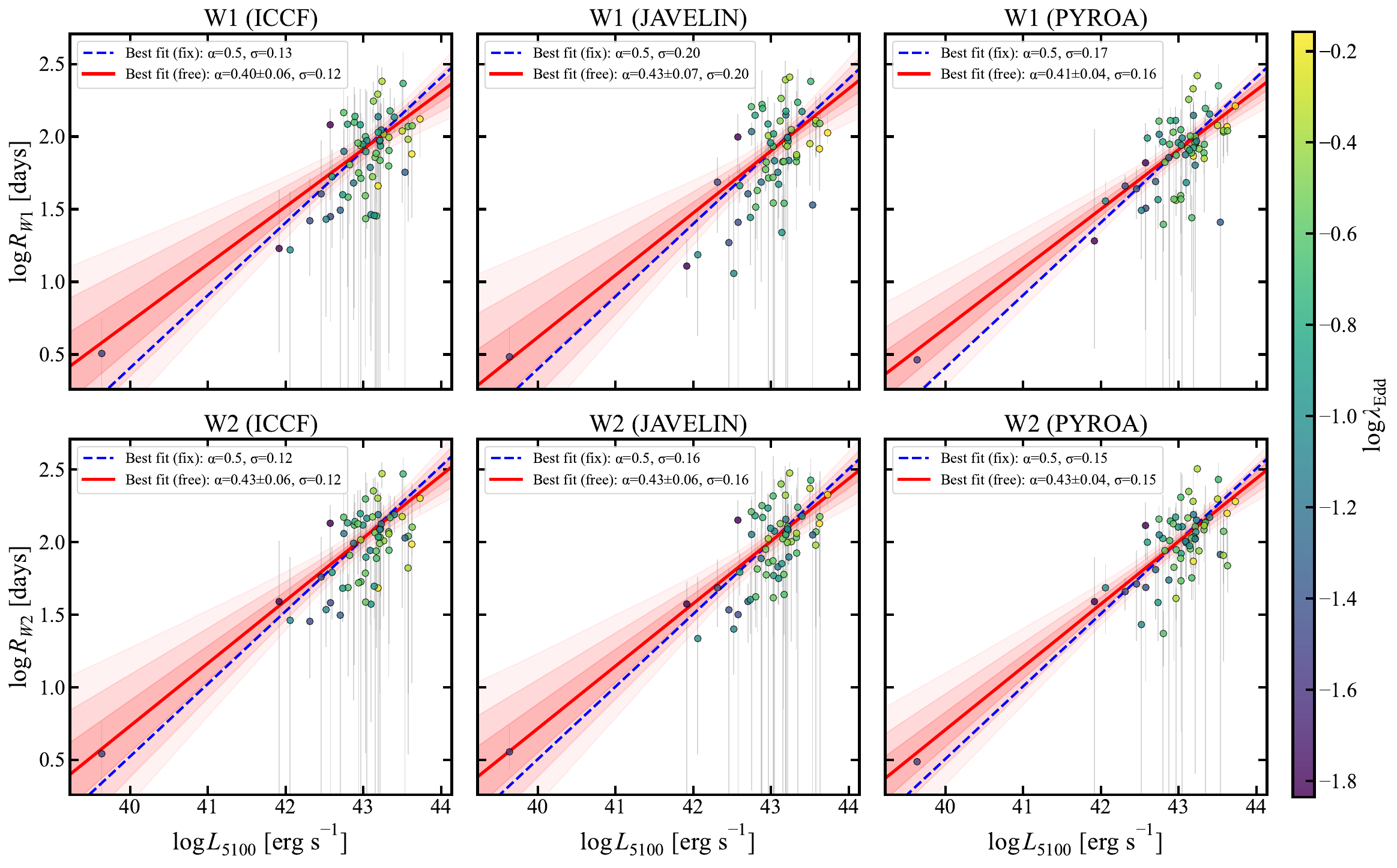}
    \end{minipage}%
    \begin{minipage}{0.5\textwidth}
        \includegraphics[width=\linewidth]{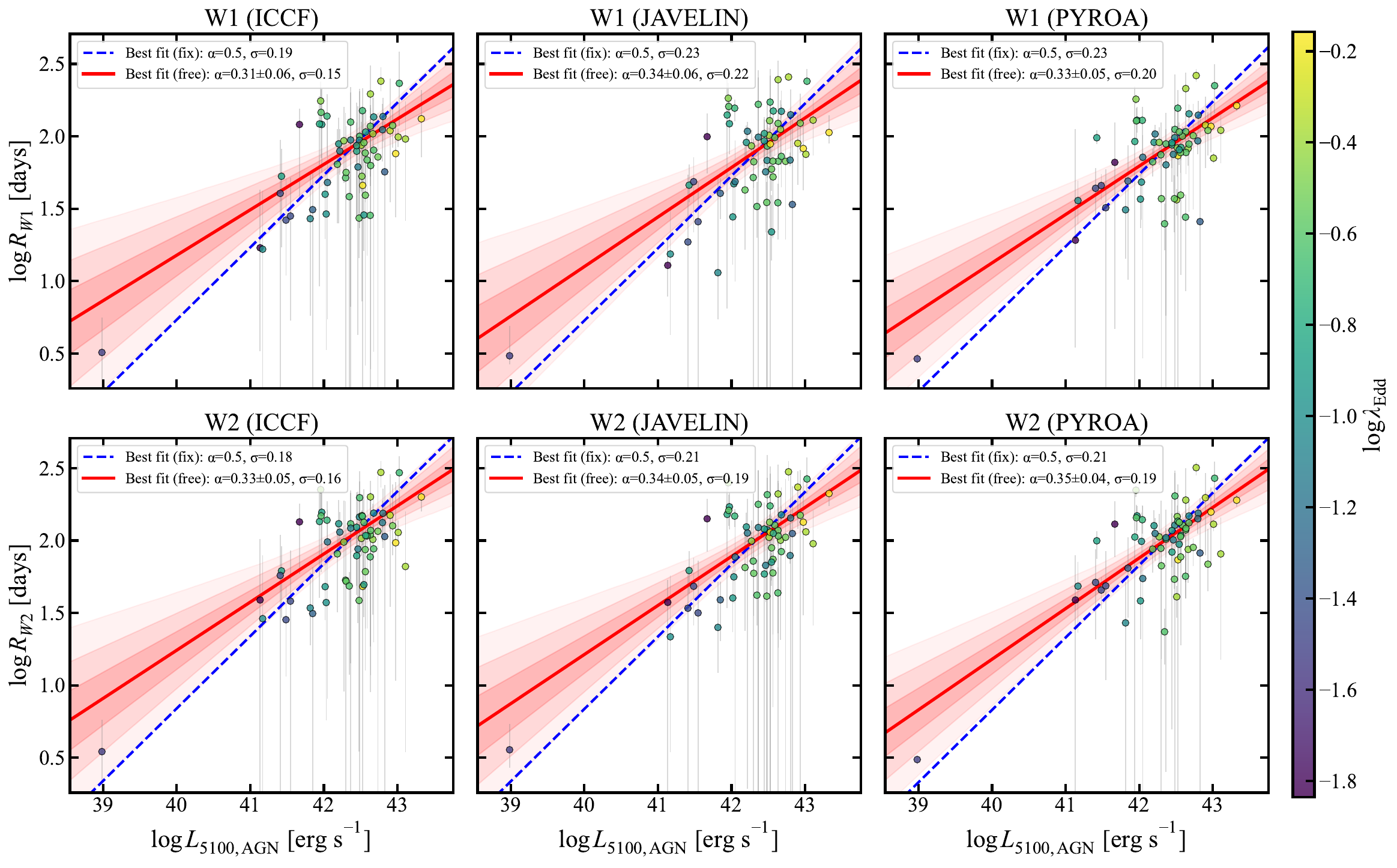}
    \end{minipage}

    \begin{minipage}{0.5\textwidth}
        \includegraphics[width=\linewidth]{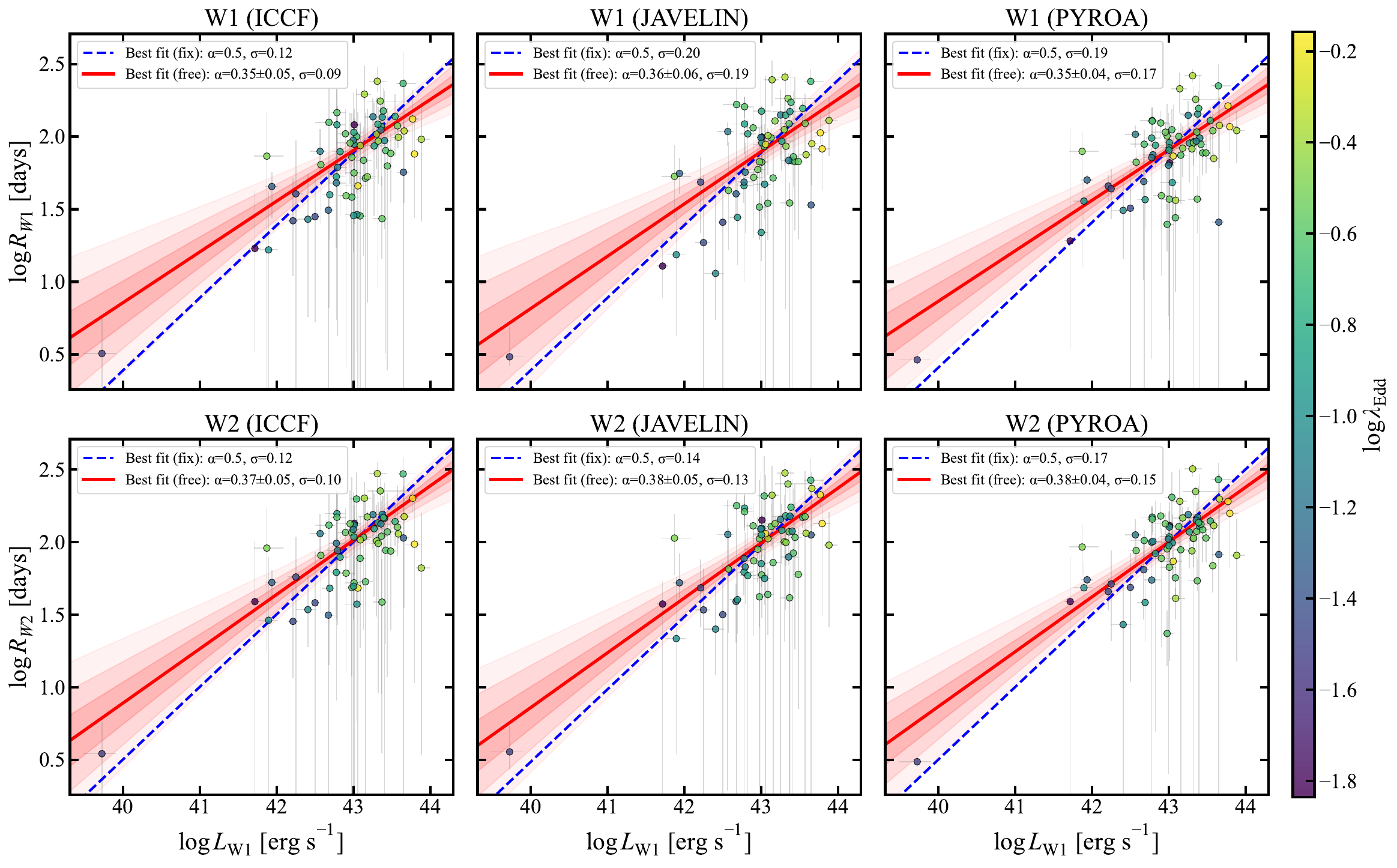}
    \end{minipage}%
    \begin{minipage}{0.5\textwidth}
        \includegraphics[width=\linewidth]{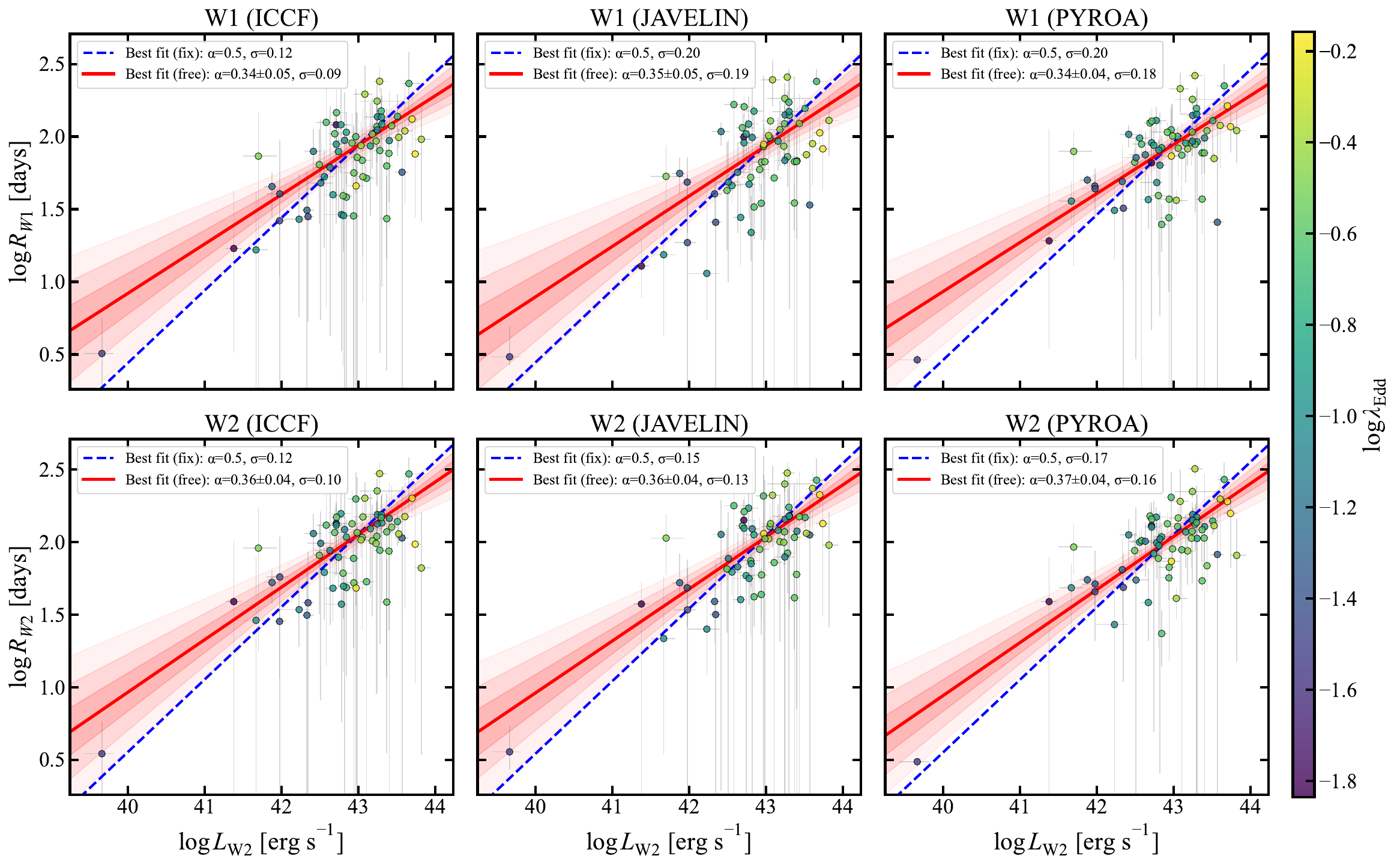}
    \end{minipage}

    \caption{Same as Figure \ref{fig:R_L_relation_logL_bol_ADC}, but the x-axis is represented by $L_{5100}$, $L_{5100,\rm AGN}$, $L_{\rm W1}$, and $L_{\rm W2}$.}
    \label{fig:R-L_relations}
\end{figure*}

\end{document}